\documentclass[twocolumn]{aastex631}

\usepackage{commath}
\usepackage{floatrow}
\usepackage{soul}
\usepackage{CJK}
\usepackage{threeparttable}
\usepackage{soul}
\usepackage{ulem}
\usepackage[caption=false]{subfig}
\usepackage{appendix}
\usepackage{xspace}

\newcommand{\msasd}{{\sc MSA-3D}\xspace}

\newcommand{\oiii}{\hbox{[O\,{\scriptsize III}]}}
\newcommand{\nii}{\hbox{[N\,{\scriptsize II}]}}
\newcommand{\sii}{\hbox{[S\,{\scriptsize II}]}}
\newcommand{\oii}{\hbox{[O\,{\scriptsize II}]}}
\newcommand{\hii}{\hbox{H\,{\scriptsize II}}}

\shorttitle{\msasd survey: metallicity gradients in $z\sim1$ galaxies}
\shortauthors{Ju et al.}
\graphicspath{{./}{figures/}}
\begin{document}
\begin{CJK*}{UTF8}{gbsn}

\title{\msasd: Metallicity Gradients in Galaxies at z $\sim$ 1 with JWST/NIRSpec Slit-stepping Spectroscopy}

\correspondingauthor{Mengting Ju, Xin Wang}
\email{jumengting@ucas.ac.cn \quad xwang@ucas.ac.cn}


\author[0000-0002-5815-2387]{Mengting Ju}
\affiliation{School of Astronomy and Space Science, University of Chinese Academy of Sciences (UCAS), Beijing 100049, China}

\author[0000-0002-9373-3865]{Xin Wang}
\affiliation{School of Astronomy and Space Science, University of Chinese Academy of Sciences (UCAS), Beijing 100049, China}
\affiliation{National Astronomical Observatories, Chinese Academy of Sciences, Beijing 100101, China}
\affiliation{Institute for Frontiers in Astronomy and Astrophysics, Beijing Normal University, Beijing 102206, China}

\author[0000-0001-5860-3419]{Tucker Jones}
\affiliation{Department of Physics and Astronomy, University of California, Davis, 1 Shields Avenue, Davis, CA 95616, USA}

\author[0000-0001-6371-6274]{Ivana Bari\v{s}i\'{c}}
\affiliation{Department of Physics and Astronomy, University of California, Davis, 1 Shields Avenue, Davis, CA 95616, USA}

\author[0000-0003-2804-0648]{Themiya Nanayakkara}
\affiliation{Centre for Astrophysics and Supercomputing, Swinburne University of Technology, Hawthorn, VIC 3122, Australia}

\author[0000-0001-9742-3138]{Kevin Bundy}
\affiliation{UCO/Lick Observatory, University of California, Santa Cruz, 1156 High Street, Santa Cruz, CA 95064, USA}

\author[0000-0002-4900-6628]{Claude-Andr{\' e} Faucher-Gigu{\` e}re}
\affiliation{Department of Physics \& Astronomy and CIERA, Northwestern University, 1800 Sherman Ave, Evanston, IL 60201, USA}

\author[0000-0002-9767-9237]{Shuai Feng}
\affiliation{College of Physics, Hebei Normal University, 20 South Erhuan Road, Shijiazhuang, 050024, China}
\affiliation{Hebei Key Laboratory of Photophysics Research and Application, 050024 Shijiazhuang, China}

\author[0000-0002-3254-9044]{Karl Glazebrook}
\affiliation{Centre for Astrophysics and Supercomputing, Swinburne University of Technology, Hawthorn, VIC 3122, Australia}

\author[0000-0002-6586-4446]{Alaina Henry}
\affiliation{Space Telescope Science Institute, 3700 San Martin Drive, Baltimore, MD 21218, USA}

\author[0000-0001-6919-1237]{Matthew A. Malkan}
\affiliation{Department of Physics and Astronomy, University of California, Los Angeles, CA 90095-1547, USA}

\author[0000-0002-1527-0762]{Danail Obreschkow}
\affiliation{International Centre for Radio Astronomy Research (ICRAR), M468, University of Western Australia, Perth, WA 6009, Australia}
\affiliation{Australian Research Council, ARC Centre of Excellence for All Sky Astrophysics in 3 Dimensions (ASTRO 3D), Australia}

\author[0000-0002-4430-8846]{Namrata Roy}
\affiliation{Center for Astrophysical Sciences, Department of Physics and Astronomy, Johns Hopkins University, Baltimore, MD, 21218}

\author[0000-0003-4792-9119]{Ryan L. Sanders}
\affiliation{Department of Physics and Astronomy, University of Kentucky, 505 Rose Street, Lexington, KY 40506, USA}
\affiliation{Department of Physics and Astronomy, University of California, Davis, One Shields Ave, Davis, CA 95616, USA}

\author[0009-0005-8170-5153]{Xunda Sun}
\affiliation{School of Astronomy and Space Science, University of Chinese Academy of Sciences (UCAS), Beijing 100049, China}

\author[0000-0002-8460-0390]{Tommaso Treu}
\affiliation{Department of Physics and Astronomy, University of California, Los Angeles, CA 90095-1547, USA}

\author[0009-0006-1255-9567]{Qianqiao Zhou}
\affiliation{School of Astronomy and Space Science, University of Chinese Academy of Sciences (UCAS), Beijing 100049, China}

\begin{abstract}

The radial gradient of gas-phase metallicity is a powerful probe of the chemical and structural evolution of star-forming galaxies, closely tied to disk formation and gas kinematics in the early universe.
We present spatially resolved chemical and dynamical properties for a sample of 25 galaxies at $0.5 \lesssim z \lesssim 1.7$ from the \msasd survey.
These innovative observations provide 3D spectroscopy of galaxies at a spatial resolution approaching JWST's diffraction limit and a high spectral resolution of $R\simeq2700$. 
The metallicity gradients measured in our galaxy sample range from $-$0.03 to 0.02 dex~kpc$^{-1}$. Most galaxies exhibit negative or flat radial gradients, indicating lower metallicity in the outskirts or uniform metallicity throughout the entire galaxy. 
We confirm a tight relationship between stellar mass and metallicity gradient at $z\sim1$ with small intrinsic scatter of 0.02 dex~kpc$^{-1}$. Our results indicate that metallicity gradients become increasingly negative as stellar mass increases, likely because the more massive galaxies tend to be more ``disky".
This relationship is consistent with the predictions from cosmological hydrodynamic zoom-in simulations with strong stellar feedback.
This work presents the effort to harness the multiplexing capability of JWST NIRSpec/MSA in slit-stepping mode to map the chemical and kinematic profiles of high-redshift galaxies in large samples and at high spatial and spectral resolution.

\end{abstract}
\keywords{galaxies: High-redshift galaxies --- galaxies: star formation --- galaxies: abundances --- galaxies: kinematics and dynamics}


\section{Introduction} \label{sec:intro}

Gas-phase metallicity is a crucial parameter for studying gas inflows and outflows. Processes such as star formation, gas accretion, and galaxy mergers can lead to local enrichment and dilution of the interstellar medium \citep[e.g,][]{Thielemann2017, Maiolino2019}. Metallicity gradients are commonly used to describe the distribution of oxygen abundance in the interstellar medium. 
The radial gradient of metallicity is a powerful probe of the chemical and structural evolution of star-forming galaxies, closely tied to their disk formation and gas kinematics. Studying these gradients, particularly in galaxies at high redshift, offers valuable information on the population evolution of galaxies \citep{Tremonti2004, Mannucci2010, Cresci2010, Queyrel2012, Swinbank2012, Jones2010_gradient, Jones2013, Troncoso2014, Ho2015, Wang2019, Wang2020, Wang2022b, Ju2022,Wangenci2023, Wangenci2024, He2024, Cheng2024, Venturi2024}.

In the local Universe, galaxies typically show a decrease in metallicity from their center to their outskirts \citep[e.g.,][]{Searle1971,  Sanchez2014, Belfiore2017, Kreckel2019, Menguiano2020}. This pattern is referred to as a ``negative metallicity gradient" \citep{Zaritsky1994, van1998}.
More recent large Integral Field Unit (IFU) surveys (e.g., CALIFA, \citealt{Sanchez2012}; MaNGA, \citealt{Bundy2015}; SAMI, \citealt{Bryant2015}) have helped establish demographic trends.
The metallicity gradients vary with galaxy morphology and stellar mass: massive disk galaxies tend to have steeper gradients, while low-mass galaxies and analogs of high-redshift systems exhibit flatter gradients \citep{Belfiore2017, Carton2018, Menguiano2018, Mingozzi2020}. However, while trends are observed in units of dex~kpc$^{-1}$,  \cite{Sanchez2014} suggest that local galaxies may have a common gradient when normalized to the effective radius ($\sim -0.1$ dex~R$_e^{-1}$, up to 2 R$_e$) in galaxies with stellar mass $\rm M_* > 10^{9.5}~M_\odot$.

Negative metallicity gradients can be explained by the inside-out growth of galaxies, where star formation initially occurs in the central regions, enriching the gas near the center first. This inside-out star formation process influences the distribution of gas-phase metallicity across galaxies \citep{Samland1997, Dave2011,Hemler2021}. In this model, metals can be redistributed by various physical processes, including stellar feedback and rotation, leading to flat gradients which can be observed in isolated high-redshift galaxies \citep{Yuan2011, Jones2013}. However, inside-out growth can not predict the positive gradients observed in some high-redshift galaxies. Indeed, there are many works suggesting that stellar feedback \citep{Gibson2013, Ma2017, Sharda2021}, mergers, and interactions \citep{Rich2012} contribute to flattening metallicity gradients. 
For example, feedback-driven winds expel metal-rich material from galaxies, which can then re-accrete in external regions, homogenizing the distribution of metals \citep{Pilkington2012, Alcazar2014,Muratov2017, Pandya2021}. 
Detailed descriptions of these processes are found in \cite{Venturi2024}.

While negative metallicity gradient slopes are well measured at $z\sim0$, it is unclear whether they persist in the early Universe.
Simulations such as IllustrisTNG \citep{Hemler2021} have shown that at $z<2$, more massive galaxies typically have flatter or even positive (also known as ``inverted'') metallicity gradients. 
In contrast, the Feedback in Realistic Environments (FIRE) simulations predict that more massive galaxies have steeper negative gradients \citep{Ma2017,Bellardini2021,Bellardini2022,Sun2024, Russell2024}. Observations of high-redshift galaxies have found that a larger fraction of these galaxies exhibit flat or inverted metallicity gradients compared to local galaxies \citep{Yuan2011,Jones2013,Wuyts2016, Leethochawalit2016, Wang2017,Wang2019, Wang2022a, Simons2021,Cheng2024}.

Notably, there is a large dispersion in metallicity gradient measurements at high redshifts across previous work, 
which may be due to large measurement errors and/or heterogeneous methods employed in the analysis. Importantly, kiloparsec-scale angular resolution is crucial to avoid biases in the spatially resolved analysis \citep{Yuan2013}, but was previously only possible with space-based slitless spectroscopy \cite[with challenges due to low spectral resolution, e.g.][]{Wang2019, Wang2020, Wang2022a} or adaptive optics (AO) assisted IFU surveys \cite[with low observing efficiency, e.g.][]{Jones2013, Leethochawalit2016}, and in both cases with a limited set of emission lines.
It is challenging to perform spatially resolved spectroscopy of high-redshift galaxies, primarily due to seeing limitations. For example, the MaNGA survey, based on a ground-based telescope, has a site seeing of approximately $1\farcs5$, corresponding to about 12 kpc at $z \sim 1$ \citep{Law2015}. This resolution exceeds the typical size of the kind of high-redshift galaxies that are thought to be Milky Way progenitors. 
Even excellent seeing of 0\farcs5 corresponds to about 4 kpc at $z\sim1$, still insufficient to resolve the effective radii of typical high-z galaxies.

Although IFS combined with AO (e.g., Keck/OSIRIS, VLT/SINFONI and ERIS, Gemini/NIFS) can achieve adequate spatial resolution, this is primarily effective at wavelengths $\lambda > 1.5 \mu$m, while instruments such as MUSE with Narrow Field Mode and laser tomography adaptive optics \citep{Bacon2010} can deliver good resolution reaching red optical wavelengths. AO observations have modest Strehl ratios in typical extragalactic fields and especially at shorter wavelengths, and are subject to beam smearing effects in all cases \citep[e.g.,][]{Burkert2016}. AO data have proven sufficient to analyze metallicity gradients with strong optical emission lines at $z\gtrsim1.5$ \citep[e.g.,][]{Yuan2011, Jones2013}, but it is hard to measure metallicities near $z = 1$ using optical emission lines such as H$\beta$, \oiii$\lambda$5007, H$\alpha$. 
While JWST offers the necessary spatial and spectroscopic resolution at the relevant wavelengths, its IFS modes are limited to observations of single objects. This makes it expensive to study metallicity gradients for large samples of galaxies.

In this work, we present measurements of metallicity gradients for 25 galaxies at $z=0.5-1.7$ using JWST/NIRSpec micro-shutter assembly (MSA) observations in a novel slit-stepping mode from the \msasd project. This new observing mode allows multiplexed IFS measurements, providing a sufficient gain in efficiency to constrain metallicity gradients for a large high-redshift galaxy sample.
By observing 43 galaxies at $z\sim1$ simultaneously, the \msasd project significantly reduces the required integration time compared to an equivalent survey with the NIRSpec IFU. In JWST Cycle 1, few projects have obtained as extensive 3D spectroscopy on galaxies at $z\sim1$ as \msasd.
Our sample is arguably the best to date at $z\sim1$ in many ways. We have (1) high angular resolution, approaching the diffraction limit of a 6.5-meter telescope; (2) exquisite S/N of nebular emission lines detected across multiple scale radii of high-$z$ galaxies; (3) far better spectral resolution than grism surveys; (4) coverage of a larger number of diagnostic emission lines as compared to most ground-based AO-assisted spectroscopy; (5) a decently large and uniform sample. For further details and description of the observations, we refer the readers to the \msasd project overview paper \citep{Ivana2024}.

The paper is organized as follows. In Section~\ref{sec:data}, we introduce the observation strategies of the slit-stepping mode, our targets, and emission lines fitting. In Section~\ref{sec:method}, we analyze the metallicity maps and the metallicity gradients of our targets. We discuss redshift and mass-dependent evolution in the metallicity gradient slopes in Section~\ref{sec:discussion}, and summarize the results in Section~\ref{summary}. We adopt the standard concordance cosmological model of $H_0=69.32\ \rm km~s^{-1}~Mpc^{-1},\ \Omega_{M} = 0.2865$.
Throughout the paper, we abbreviate the forbidden lines with 
$\oiii\lambda\lambda4959,5007=  \oiii$,
$\nii\lambda\lambda6548,6584= \nii$,
$\sii\lambda\lambda6717,6731= \sii$, 
if presented without wavelength values.

\section{Data} \label{sec:data}
\subsection{Sample and observations}

\begin{figure*}
    \centering
    \includegraphics[width=1.0\textwidth,clip,trim={0 0 0 0}]{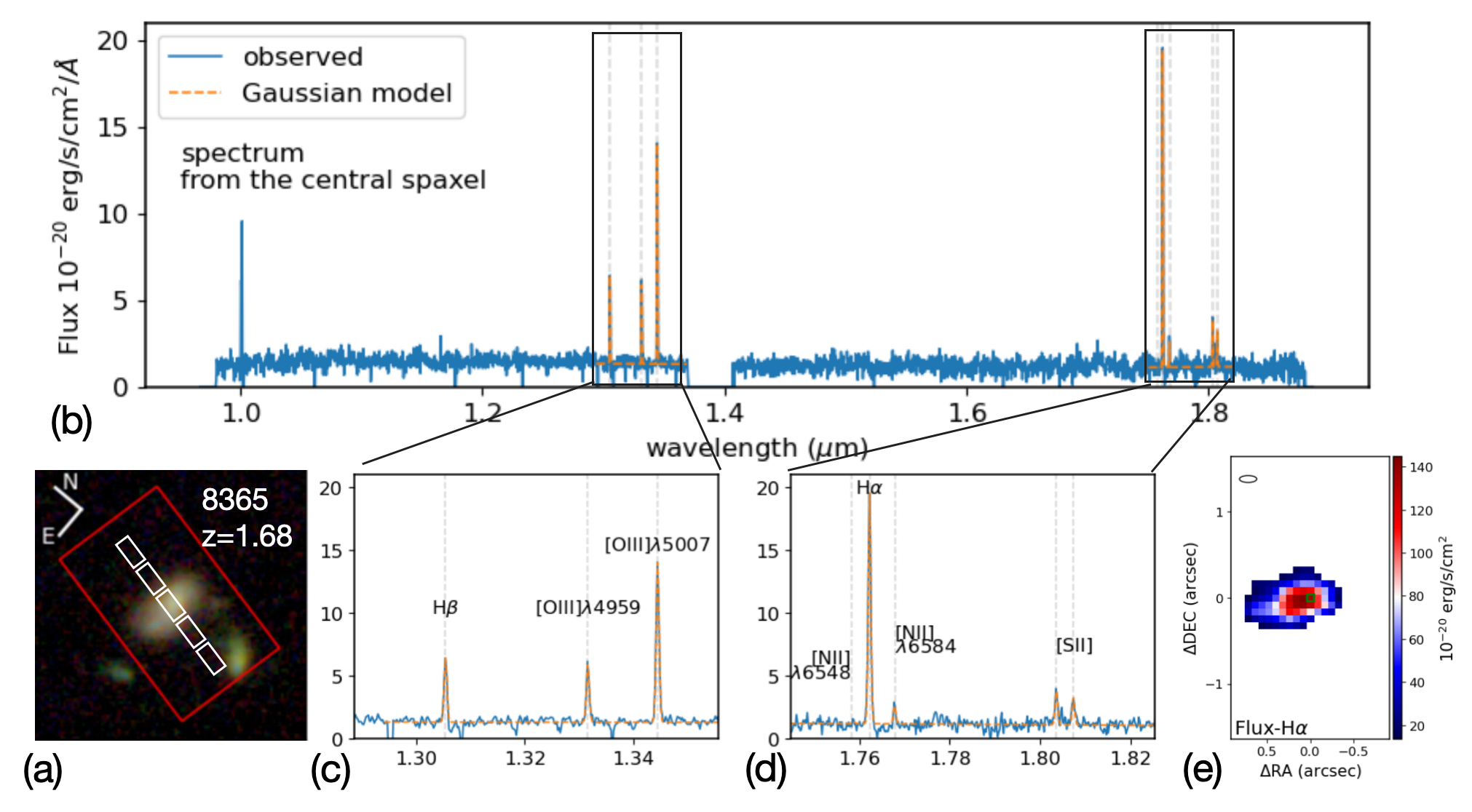} 
    \caption{An example of the observed spectrum from the spaxel in the galaxy center and Gaussian models of the observed emission lines. Panel (a): The color-composite image of galaxy ID8365 is made from the JWST/NIRCam imaging (blue: F115W, green: F277W, red: F444W). The red box ($1\farcs8 \times 3\farcs0$) represents the entire field of view of our slit-stepping 3D spectroscopy and the white boxes mark the relative sizes of the 5 open microshutter slitlets. The observing strategy of the \msasd project acquires spectroscopy of our targets in 63 pointings in total,     
    moving slitlets in 9 steps of 0\farcs2 in the dispersion direction and 7 dithers of 0\farcs075 in the cross-dispersion direction. For further details, refer to \cite{Ivana2024}. Panel (b) shows the full observed spectrum from the spaxel in the galaxy center indicated as a green box in panel (e), with the orange curves being the Gaussian-fitted emission line models. We detect pronounced nebular emission features in the galaxy center.
    We also zoom-in on the line fitting results in panels (c) and (d). The grey vertical dashed lines represent the locations of H$\beta$, $\oiii$, H$\alpha$, $\nii$, and $\sii$. Panel (e): the flux map of the H$\alpha$ emission line, where the ellipse represents the resolution element. }

    \label{fig:image}
\end{figure*}

The \msasd project obtained its first observations from March 29 to March 30, 2023 (JWST Cycle 1, GO-2136, PI: Jones).
Using JWST/NIRSpec, we observed 43 star-forming galaxies at $z=0.5-1.7$ in the Extended Groth Strip (EGS) field using a slit-stepping mode to obtain multiplexed integral field spectroscopy (IFS). The field contains extensive photometric and spectral survey data. 
We selected galaxies with reliable redshifts from the CANDELS survey \citep{Koekemoer2011} and the 3D-HST survey catalogue \citep{Skelton2014, Momcheva2016},
ensuring that emission lines such as H$\alpha$, \nii, \oiii, and H$\beta$ are observable using the G140H/F100LP grating/filter. Properties of the 43 targeted galaxies are listed in Table~\ref{table:1}. Most of the targets are located on the ``star-forming main sequence," which is characterized by a specific star formation rate of approximately 10$^{-9}$ yr$^{-1}$ at $z\sim1$ \citep[e.g.,][]{Whitaker2014}. The NIRSpec grating and filter, G140H/F100LP, covered a wavelength range of $0.97\mu m \leq \lambda \leq 1.82 \mu m$ with a spectral resolution of $R \sim 2700$ to map strong nebular emission lines. We employed a consistent MSA mask setup across 63 distinct pointings, moving one slit width ($0\farcs2$) along the dispersion axis in each of the 9 steps, and one barshadow width ($0\farcs075$) along the cross-dispersion axis in each of the 7 steps to avoid the influence of the bar between all shutters. The total exposure time was 20.4 hours and the effective integration time at each spatial position was 117 minutes. Detailed observing strategies and data processing methods are discussed in the \msasd project overview paper \citep{Ivana2024}. The FoV in the datacubes spans $1\farcs8\times(2\farcs0-3\farcs0)$, depending on the number of slitlets (3-5) used to observe each galaxy. The resolution element in a typical reconstructed data cube is about $\rm 0\farcs20 \times 0\farcs08\ (\sim\ 1.63 \times\ 0.65\ kpc\ at\ z=1)$. In this paper, we analyze data cubes that were interpolated onto a new grid with a spatial sampling of 0\farcs08 $\times$ 0\farcs08.

We present an example of the observed spectrum from the central spaxel of galaxy ID8365 ($z=1.68$) in Figure~\ref{fig:image}. The stellar mass of this galaxy is $\log(M_{*}/M_{\odot}) = 9.79$ and its SFR is 9.76 $\rm M_\odot\ yr^{-1}$. These values were obtained from the UVCANDELS catalog, and are reported in Table \ref{table:1}. The properties were estimated using Dense Basis SED modeling based on HST observations ranging from F275W to F160W filters \citep{Sun2023, Wang2024RNAAS, Mehta2024}. In panel (a), we present a three-color image obtained from JWST/NIRCam imaging using the F115W, F277W, and F444W filters. For this galaxy, we selected five slitlets for observation using the slit-stepping method (indicated by white rectangles). 
The red box represents the FoV of the resulting pseudo-IFU datacube, which is $1\farcs8 \times\ 3\farcs0$. The chip gap occurs near 1.4 $\mu$m (corresponding to a rest wavelength of $\sim\ 5223$ \AA\ for this target). The spectrum exhibits very distinct emission lines, particularly strong ones like the $\oii\lambda\lambda3727,3729$, H$\beta$, $\oiii$, H$\alpha$, $\nii$, $\sii$ emission line. 
The \msasd data provide excellent S/N of these emission lines in individual spaxels.

\begin{table*}[]
  \scriptsize
  \caption{Measured properties for the parent sample of 43 galaxies on which we perform slit-stepping 3D spectroscopy in the \msasd project.}
  \tabcolsep=0.08cm
  \label{table:1}
  \centering
\begin{tabular}{cccccccccc}
\hline\hline
ID	&	 RA	&	 Dec		&	redshift$^a$	&	stellar mass$^a$	&	SFR$^a$ & $\log \left({\rm \frac{sSFR}{yr^{-1}}} \right)$ &  emission lines$^b$ &	\multicolumn{2}{c}{metallicity gradient and 1-$\sigma$ uncertainty}  \\
-	&	 Degree	&	 Degrees		&	-	& $\rm log(M_{*}/M_{\odot})$	& $\rm M_{\odot} yr^{-1}$ & - &  - & $\rm dex\ kpc^{-1}$  & $\rm dex\ R_e^{-1}$ \\
\hline
2111	&	215.0627824	&	52.9070766	&	0.58 	&	10.37 	&	0.33 	&	-10.85	& 	H$\alpha$, \nii, \sii	& 		-		& 		-		\\
2145	&	215.0694675	&	52.9108516	&	1.17 	&	9.44 	&	3.00 	&	-8.96	& 	H$\alpha$, \nii, \sii, H$\beta$, \oiii	& 	0.0015 	$\pm$	0.0220 	& 	0.0917 	$\pm$	0.1016 	\\
2465	&	215.0704381	&	52.9137241	&	1.25 	&	9.70 	&	1.22 	&	-9.61	& 	H$\alpha$, \nii, \sii, H$\beta$, \oiii	& 		-		& 		-		\\
2824	&	215.0685025	&	52.914333	&	0.98 	&	9.92 	&	0.39 	&	-10.33	& 	H$\alpha$, \nii, \sii, \oiii	& 	0.0211 	$\pm$	0.0247 	& 	0.0862 	$\pm$	0.0867 	\\
3399	&	215.042511	&	52.8996031	&	1.34 	&	10.25 	&	7.25 	&	-9.39	& 	H$\alpha$, \nii, \sii, H$\beta$, \oiii	& 		-		& 		-		\\
4391	&	215.0676175	&	52.9232437	&	1.08 	&	9.83 	&	4.73 	&	-9.16	& 	H$\alpha$, \nii, H$\beta$, \oiii	& 	-0.0013 	$\pm$	0.0117 	& 	-0.1062 	$\pm$	0.0433 	\\
6199	&	215.0450158	&	52.9194652	&	1.59 	&	9.92 	&	35.44 	&	-8.37	& 	H$\alpha$, \nii, \sii, H$\beta$	& 	-0.0214 	$\pm$	0.0072 	& 	-0.1152 	$\pm$	0.0544 	\\
6430	&	215.0131444	&	52.8980378	&	1.17 	&	10.23 	&	7.52 	&	-9.35	& 	H$\alpha$, \nii, \sii, H$\beta$, \oiii$\lambda$4959	& 	0.0081 	$\pm$	0.0062 	& 	-0.0162 	$\pm$	0.0342 	\\
6848	&	215.0355588	&	52.9166925	&	1.57 	&	8.68 	&	0.03 	&	-10.2	& 	H$\alpha$, \nii, \sii, \oiii	& 		-		& 		-		\\
7314	&	214.9989097	&	52.8925151	&	1.28 	&	9.66 	&	4.79 	&	-8.98	& 	H$\alpha$, \nii, H$\beta$, \oiii 	& 	-0.0162 	$\pm$	0.0170 	& 	-0.0080 	$\pm$	0.0891 	\\
7561	&	215.0609094	&	52.9383828	&	1.03 	&	9.26 	&	2.95 	&	-8.79	& 	H$\alpha$, \nii, \sii, H$\beta$, \oiii	& 		-		& 		-		\\
8365	&	215.0599904	&	52.9422373	&	1.68 	&	9.79 	&	9.76 	&	-8.8	& 	H$\alpha$, \nii, \sii, H$\beta$, \oiii	& 	0.0059 	$\pm$	0.0384 	& 	-0.0027 	$\pm$	0.0948 	\\
8512	&	215.0497806	&	52.9380795	&	1.10 	&	10.56 	&	9.30 	&	-9.59	& 	H$\alpha$, \nii, \sii, H$\beta$, \oiii	& 	-0.0241 	$\pm$	0.0053 	& 	-0.1650 	$\pm$	0.0450 	\\
8576	&	215.0595679	&	52.9434335	&	1.57 	&	9.80 	&	25.67 	&	-8.39	& 	H$\alpha$, \nii, \sii, H$\beta$, \oiii	& 	-0.0313 	$\pm$	0.0142 	& 	-0.0756 	$\pm$	0.0475 	\\
8942	&	215.0094032	&	52.9100655	&	1.18 	&	10.36 	&	6.51 	&	-9.55	& 	H$\alpha$, \nii, \sii, H$\beta$, \oiii$\lambda$4959	& 	0.0046 	$\pm$	0.0160 	& 	0.0079 	$\pm$	0.0384 	\\
9337	&	214.9957065	&	52.9019407	&	1.17 	&	9.41 	&	3.91 	&	-8.82	& 	H$\alpha$, \nii, \sii, H$\beta$, \oiii	& 		-		& 		-		\\
9424	&	214.9926533	&	52.900911	&	0.98 	&	10.04 	&	5.50 	&	-9.3	& 	H$\alpha$, \nii, \sii	& 	0.0085 	$\pm$	0.0056 	& 	0.0361 	$\pm$	0.0301 	\\
9482	&	215.0530158	&	52.9442406	&	1.21 	&	10.21 	&	0.12 	&	-11.13	& 	H$\alpha$, \nii \sii, H$\beta$, \oiii	& 		-		& 		-		\\
9527	&	215.0085005	&	52.9123786	&	1.42 	&	10.32 	&	9.09 	&	-9.36	& 	H$\alpha$, \nii, H$\beta$, \oiii	& 		-		& 		-		\\
9636	&	215.0365985	&	52.9328783	&	0.74 	&	9.87 	&	0.51 	&	-10.16	& 	H$\alpha$, \nii, \sii	& 	-0.0167 	$\pm$	0.0221 	& 	-0.0545 	$\pm$	0.1426 	\\
9812	&	215.0402843	&	52.9376004	&	0.74 	&	10.37 	&	2.93 	&	-9.9	& 	H$\alpha$, \nii, \sii	& 	-0.0336 	$\pm$	0.0051 	& 	-0.1715 	$\pm$	0.0312 	\\
9960	&	215.031894	&	52.9331513	&	1.51 	&	11.20 	&	11.76 	&	-10.13	& 	H$\alpha$, \nii, \sii	& 	-0.0274 	$\pm$	0.0116 	& 	-0.1707 	$\pm$	0.0750 	\\
10107	&	214.9817759	&	52.8975743	&	1.01 	&	10.44 	&	3.25 	&	-9.93	& 	H$\alpha$, \nii, \sii, \oiii	& 	-0.0238 	$\pm$	0.0083 	& 	-0.0657 	$\pm$	0.0308 	\\
10502	&	214.9857711	&	52.9033048	&	1.23 	&	10.55 	&	9.17 	&	-9.59	& 	H$\alpha$, \nii, H$\beta$, \oiii	& 		-		& 		-		\\
10752	&	215.0403761	&	52.9413773	&	1.73 	&	9.67 	&	15.61 	&	-8.48	& 	H$\beta$, \oiii	& 		-		& 		-		\\
10863	&	215.0551264	&	52.9529908	&	1.03 	&	9.43 	&	4.97 	&	-8.73	& 	H$\alpha$, \nii, H$\beta$, \oiii	& 		-		& 		-		\\
10910	&	215.0561932	&	52.9553701	&	0.74 	&	10.06 	&	0.93 	&	-10.09	& 	H$\alpha$, \nii, \sii	& 	-0.0113 	$\pm$	0.0066 	& 	-0.0328 	$\pm$	0.0406 	\\
11225	&	215.0415788	&	52.9454655	&	1.05 	&	10.01 	&	6.83 	&	-9.18	& 	H$\alpha$, \nii, \sii, H$\beta$, \oiii	& 	0.0068 	$\pm$	0.0071 	& 	-0.0275 	$\pm$	0.0486 	\\
11539	&	214.9819651	&	52.9051336	&	1.61 	&	10.72 	&	29.68 	&	-9.25	& 	H$\beta$, \oiii	& 		-		& 		-		\\
11702	&	214.9794086	&	52.9031601	&	1.23 	&	9.77 	&	4.19 	&	-9.15	& 	H$\beta$, \oiii	& 		-		& 		-		\\
11843	&	215.0390468	&	52.9471037	&	1.46 	&	11.09 	&	2.92 	&	-10.62	& 	H$\alpha$, \nii, \sii, H$\beta$, \oiii	& 		-		& 		-		\\
11944	&	215.0369901	&	52.9453937	&	1.04 	&	9.62 	&	3.46 	&	-9.08	& 	H$\alpha$, \nii, \sii, H$\beta$, \oiii	& 	-0.0157 	$\pm$	0.0090 	& 	-0.0430 	$\pm$	0.0619 	\\
12015	&	215.0323151	&	52.9431798	&	1.24 	&	10.54 	&	6.36 	&	-9.74	& 	H$\alpha$, \nii, \sii, H$\beta$, \oiii	& 	-0.0302 	$\pm$	0.0126 	& 	-0.0940 	$\pm$	0.0437 	\\
12071	&	215.0219665	&	52.9360567	&	1.28 	&	9.87 	&	2.76 	&	-9.43	& 	H$\alpha$, \nii, \sii, H$\beta$, \oiii	& 		-		& 		-		\\
12239	&	215.0495343	&	52.9560252	&	0.89 	&	9.78 	&	0.10 	&	-10.78	& 	H$\alpha$, \nii, \sii	& 	-0.0114 	$\pm$	0.0215 	& 	-0.0409 	$\pm$	0.0537 	\\
12253	&	215.0442096	&	52.9520815	&	1.03 	&	9.22 	&	1.37 	&	-9.08	& 	\sii, H$\beta$, \oiii	& 		-		& 		-		\\
12773	&	215.029765	&	52.9451588	&	0.95 	&	9.67 	&	2.27 	&	-9.31	& 	H$\alpha$, \nii, \sii	& 	-0.0001 	$\pm$	0.0144 	& 	0.0187 	$\pm$	0.0510 	\\
13182	&	214.99983	&	52.9268186	&	1.54 	&	10.47 	&	100.35 	&	-8.47	& 	H$\beta$, \oiii	& 		-		& 		-		\\
13416	&	215.0252815	&	52.9456859	&	1.54 	&	10.11 	&	51.46 	&	-8.4	& 	H$\alpha$, \nii, \sii, \oiii	& 	-0.0020 	$\pm$	0.0145 	& 	-0.0189 	$\pm$	0.0619 	\\
18188	&	214.9839579	&	52.9413562	&	0.82 	&	9.81 	&	0.83 	&	-9.89	& 	H$\alpha$, \nii, \sii	& 	-0.0125 	$\pm$	0.0095 	& 	-0.0319 	$\pm$	0.0529 	\\
18586	&	214.9712282	&	52.9337862	&	0.76 	&	9.61 	&	1.17 	&	-9.54	& 	H$\alpha$, \nii, \sii	& 	-0.0035 	$\pm$	0.0225 	& 	-0.0086 	$\pm$	0.0622 	\\
19382	&	214.9765628	&	52.9414977	&	1.03 	&	9.57 	&	1.16 	&	-9.51	& 	H$\alpha$, \nii, \sii, \oiii	& 		-		& 		-		\\
29470	&	214.9689505	&	52.9453733	&	1.04 	&	9.93 	&	7.95 	&	-9.03	& 	H$\alpha$, \nii, \sii, H$\beta$, \oiii	& 	-0.0244 	$\pm$	0.0066 	& 	-0.1239 	$\pm$	0.0478 	\\
\hline
\end{tabular}\\

Notes:
$^a$ The stellar population properties presented here are obtained from the UVCANDELS catalog.\\
$^b$ Here we list the strong nebular emission lines fitted in this work. The forbidden lines correspond to both doublets unless otherwise specified.
\end{table*}

\subsection{Emission line fitting}

In this work, our primary focus is on the metallicity gradients, which are derived from the ratios of the emission line fluxes. Our sources are star-forming galaxies, with strong emission lines (as shown in Figure~\ref{fig:image}).
We fit Gaussian profiles to various emission lines to determine their line flux, velocity, and velocity dispersion. Specifically, a three-parameter Gaussian curve was fitted to the following emission lines for each spaxel: H$\beta$, \oiii, H$\alpha$, \nii, \sii. In many cases, not all of the lines are observed, depending on the source redshift and wavelength coverage. 
Table~\ref{table:1} lists the emission lines observed for each galaxy.

We typically do not detect Balmer absorption or other stellar features in the spectra of individual spaxels. Therefore, we fit the spectra using a straight line to model the continuum, plus Gaussian profiles for the emission lines. Based on spectral energy distribution modeling of our targets, we find that stellar Balmer absorption has little effect on the results \citep[e.g.,][]{Zahid2011}, causing the metallicity to be overestimated by $\lesssim 0.05$ dex with negligible effect on the gradient slopes.
Each spectrum was fitted in two regions with rest-frame wavelengths near 4820-5100 \AA\ (including H$\beta$ and \oiii) and 6520-6800 \AA\ (including H$\alpha$, \nii, and \sii). We fit the emission lines plus continuum simultaneously.
All lines are fit with the same redshift (i.e., velocity) and width (i.e., dispersion), using the values measured for H$\alpha$.
To calculate intrinsic line widths, we correct for the instrument resolution by subtracting it in quadrature from the best-fit value.
The resolution curve for the G140H grating shows that the average instrumental FWHM across the emission line wavelength range is approximately 5.2 \AA, corresponding to an instrumental dispersion of roughly 50~km~s$^{-1}$.

In Figure~\ref{fig:image}, we show an example of Gaussian emission line fits with orange lines in panels (c) and (d). The emission lines are distinct in the spaxel of galaxy ID8365, yielding a high S/N for the line flux measurements. For instance, the flux of H$\alpha$ is $\rm (133.83 \pm 2.28 )\times10^{-20}~erg/s/cm^2$ and the equivalent width (EW) is $\rm (44.40 \pm 1.57 ) \AA$ in the rest frame. The uncertainties in the flux and EWs are calculated based on the residuals between the best-fit linear and Gaussian models and the observed spectra. This statistical uncertainty represents the emission S/N, and does not include additional sources of uncertainty such as underlying stellar absorption.
The flux map of the H$\alpha$ emission line for galaxy ID8365 is shown in panel (e). The ellipse indicates the resolution element ($\rm 0\farcs20 \times 0\farcs08$).

The sky coverage of the IFS data cubes is larger than the effective radii of our targets (Figure~\ref{fig:image}), and in general, there are many spaxels with no detectable emission.
Therefore, it is necessary to select the reliable spaxels associated with each galaxy for analysis. We select the spaxels from the target sources using the properties of H$\alpha$, requiring the S/N of the H$\alpha$ flux to be greater than 10 and the S/N of the H$\alpha$ EW to be greater than 5. 
The EW S/N $>$ 5 requirement is typically not a limiting factor, as we exclude only 10\% of spaxels which otherwise meet the emission line flux requirement.
In addition, we exclude unreliable spaxels by visually inspecting the 3-color images and spectra. Consequently, we obtain spaxels from the target galaxies that are suitable for analysis.
The resulting maps of all galaxies analyzed in this paper, including their H$\alpha$ flux, observed velocity, and observed velocity dispersion, are shown in Figure~\ref{fig:1} in Appendix~\ref{sec:appendix}.

\section{Gas-phase metallicity and its radial gradient}\label{sec:method}

\subsection{Measuring chemical abundances}

Most metallicity diagnostics are calibrated on \hii\ regions. Therefore, before obtaining gas-phase metallicity maps, we need to verify that the emission lines originate from star-forming regions.
Isolating individual \hii\ regions is highly challenging in the high-$z$ universe, due to limited resolution \citep{Sanchez2014}.
In this work, we exclude regions not classified as star-forming in the BPT diagram \citep{BPT1981,kewley2001,kauffmann2003a} using the \nii/H$\alpha$ versus \oiii/H$\beta$ ratios. However, because of detector gaps and limited wavelength ranges, not all four emission lines are detected in each galaxy (Table~\ref{table:1}). \cite{whan2011} demonstrated a tight correlation between \nii/H$\alpha$ and EW(H$\alpha$), indicating that pure star-forming regions have EW(H$\alpha$) $> 3$~\AA\ and log(\nii/H$\alpha$) $< -0.4$. 
For galaxies where \hii\ regions cannot be identified using the BPT diagram, we instead use the WHAN diagram of EW(H$\alpha$) and log(\nii/H$\alpha$).
Figure~\ref{fig:1} in Appendix~\ref{sec:appendix} presents the BPT or WHAN diagrams and their maps color-coded according to the positions of individual regions for galaxies in our sample.
We compared the results of WHAN and BPT diagrams for the 10 galaxies where both diagnostics are available. In both diagrams, spaxels classified as non-star-forming are largely in the outer regions of low S/N and constitute a small fraction of the total spaxels (Figure~\ref{fig:1}). Additionally, the majority of spaxels classified as star-forming by the WHAN diagram were also classified as star-forming in the BPT diagram, indicating strong consistency between these diagnostic methods. This analysis indicates that WHAN diagrams can be effectively utilized for our sample to provide consistent results in cases where the BPT diagram is not available.

Various strong-line methods use different indicators to measure gas-phase metallicity. Examples include $\rm O3N2 = \log\frac{\oiii\lambda5007/H\beta}{\nii\lambda6584/H\alpha}$ \citep{pp04, marino2014}, $\rm N2 = \log\frac{\nii\lambda6584}{H\alpha}$ \citep{Storchi1994, Denicolo2002, marino2014}, and the N2S2H$\alpha$ diagnostic \citep[using H$\alpha$, \nii, and \sii;][]{Dopita2016,Cameron2019}. O3N2 and N2 are among the most widely used indicators for calculating gas-phase metallicity. 
These indicators are minimally affected by dust extinction since they use pairs of emission lines which are close in wavelength. 
Given the emission line coverage of our targets (Table \ref{table:1}), the O3N2 indicator can be used for 20 galaxies, the N2 indicator for 38 galaxies, and N2S2H$\alpha$ for 33 galaxies. 
While N2S2H$\alpha$ has some advantages over N2, it requires better S/N such that there are fewer spaxels with reliable measurements.
Thus, for this work, we use the N2 indicator to give the largest homogeneous sample. 
While we use both WHAN and BPT diagrams to select star-forming spaxels, we have verified that these give consistent metallicity gradients for the 10 galaxies where both diagnostics are available.
It is important to use the same calibration for all galaxies in our analysis in order to avoid systematic differences \citep[e.g.,][]{Kewley2008, Sanchez2012b, Peimbert2012}. 
The N2 values for our targets range from -2.5 to -0.3, and we use the fit given by \cite{pp04} which is valid across this range:
\begin{equation}
\begin{split}
   &12+\mathrm{log(O/H)}=\\
   &9.37+2.03\times \mathrm{N2}+1.26\times (\mathrm{N2})^2+0.32\times (\mathrm{N2})^3 
   \label{eq:n2}
\end{split}
\end{equation} 
with a systematic 1-$\sigma$ uncertainty of 0.18 dex. We thus combine this scatter in quadrature when deriving metallicities using this strong line calibration.

We also visually inspect the resulting metallicity maps of these galaxies. Spaxels with uncertainty of $> 0.25$ dex in metallicity were discarded before deriving the metallicity gradient. Discarding these spaxels has no significant effect on the results.
To ensure sufficient spatial coverage to measure a reliable gas-phase metallicity gradient, we required galaxies to have a minimum of 20 spaxels located in the star-forming regions of the BPT or WHAN diagram.
This requirement results in 25 galaxies suitable for metallicity gradient measurements (out of 38 galaxies with suitable wavelength coverage). Of these, 10 galaxies relied on the BPT diagram to determine their star-forming spaxels, while the remaining galaxies utilized the WHAN diagram.
The metallicity maps for each galaxy are shown in Figure~\ref{fig:1} in Appendix~\ref{sec:appendix}.

\subsection{Deriving abundance gradients}

\begin{figure}
    \centering
    \includegraphics[width=0.95\textwidth,clip,trim={0 0 0 0}]{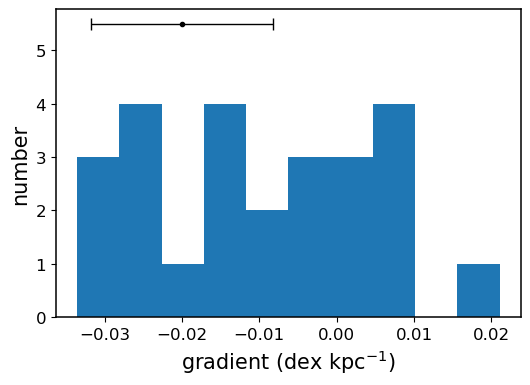} 
    \caption{The distribution of metallicity gradient slopes, in units of dex~kpc$^{-1}$. The black error bar shows the median 1-$\sigma$ uncertainty, 
    highlighting the very small measurement uncertainties afforded by the high-quality \msasd data set.
    The metallicity gradients of our 25 galaxies are predominantly negative or flat, with slopes typically more positive than those of local spiral galaxies.}    
    \label{fig:thiswork}
\end{figure}

In order to measure the de-projected distance (R/kpc) of each spaxel from the center, it is essential to determine the galaxy center, position angle, and axis ratio (b/a). These structural parameters were measured using multi-wavelength HST photometry from the CANDELS survey \citep{Koekemoer2011}. We used the position angle and b/a values obtained from the CANDELS catalog \citep{Stefanon2017}.
The centers of the galaxies are determined as the peaks of the continuum maps from the data cubes, using rest wavelengths around $5500 \pm 25$~\AA\ whenever possible. If this is not feasible, wavelengths around $6500 \pm 25$~\AA\ are used. Both regions represent optical continuum which is free of strong emission lines.
The metallicity gradient (i.e., O/H as a function of de-projected radius) for each galaxy is shown in Figure~\ref{fig:1} in Appendix~\ref{sec:appendix}. We also adopt the effective radius $\rm R_e$ as reported in \cite{Stefanon2017} to examine metallicity as a function of normalized distance $\rm R/R_e$ for each galaxy. 
Figure~\ref{fig:1} includes both physical (R/kpc) and normalized ($\rm R/R_e$) radii, on the lower and upper axes respectively.
We perform a linear fit to the gradients from 0.5 kpc to the edge of the field of view. This fit is shown by the red dashed line, with the slope of the line indicated in the subplot. 
The central region is not used in order to mitigate beam smearing effects. 
We additionally fit a linear model within the range 0.5 -- 2.5 ${\rm R_e}$, following a similar approach to that used in \cite{Belfiore2017} for purposes of comparison. The distribution of gradients in physical (at $>0.5$~kpc) units is shown in Figure~\ref{fig:thiswork}. The gradient slopes measured in our sample range from $-$0.03 to 0.02 dex~kpc$^{-1}$, with a median uncertainty of 0.01 dex~kpc$^{-1}$. We note that the errors of our gradient measurements are primarily dominated by the scatter of 0.18 dex from the metallicity inference based on the \citet{pp04} strong line calibration. 
Most galaxies exhibit negative or flat radial gradients, indicating that metallicity is lower or similar in the outskirts compared to the centers.
However, the slopes are typically less negative than local galaxies, suggesting that flatter or even inverted metallicity gradients are more common at $z\sim1$ compared to $z\sim0$ (see Section~\ref{sec:discussion}).

The slopes of gradients can be misleading when radial profiles are not linear. Therefore, it is more accurate to examine the complete profiles \citep[e.g.,][]{Oyarzun2019}. In the subplots of Figure~\ref{fig:1}, we also show the metallicity profiles represented by the average values in radial bins. In observing the profiles and slopes of these galaxies, we identified some interesting phenomena. The best-fit linear slope alone is insufficient to fully characterize the metallicity gradient in some galaxies. For instance, although the slopes of galaxies ID4391 and ID13416 are similar, their profiles differ significantly. These distinct profiles may indicate different galaxy formation processes, with ID13416 likely undergoing a more quiescent state than ID4391. In this paper, we attempt to quantify these profiles by calculating the root-mean-square error (RMSE), which is the square root of the mean of the differences between the observed metallicity in spaxels and predicted values provided by the linear models. The RMSE scatter in our sample ranges from 0.04 to 0.11 dex~kpc$^{-1}$. The majority display relatively low scatter, such that the radial metallicity profile is well approximated with a linear fit.
Only 4 out of 25 galaxies have metallicity gradients associated with large RMSE scatter ($\rm >0.1\ dex\ kpc^{-1}$).


\section{Discussion} \label{sec:discussion}

\subsection{Metallicity gradients versus redshift and stellar mass}

\begin{figure*}
    \centering
    \includegraphics[width=0.95\textwidth,clip,trim={0 0 0 0}]{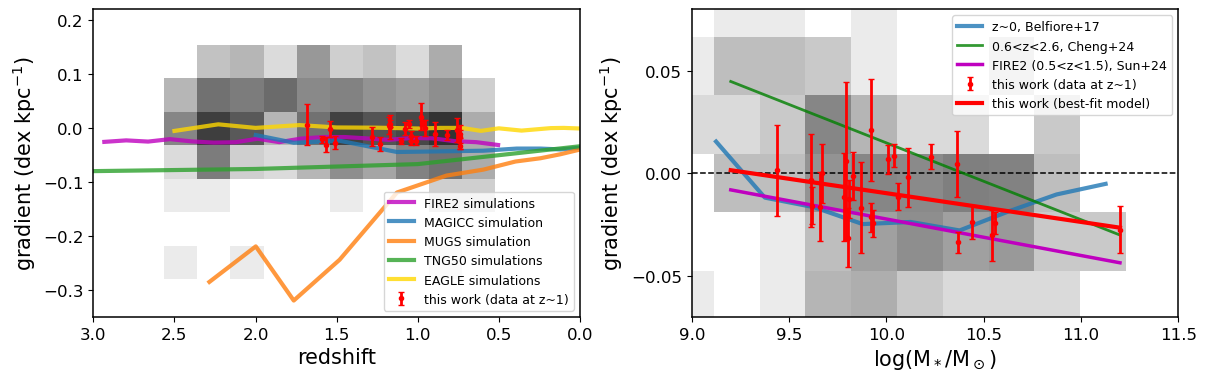}\\
    \caption{Correlations among the metallicity gradient, redshift, and stellar mass. Left: Metallicity gradients as a function of redshift. The red points with 1-$\sigma$ error bars correspond to our results. The orange and blue lines correspond to the MUGS (``weak'' feedback) and MAGICC (``enhanced'' feedback) simulations from \cite{Gibson2013}, respectively. The green line represents the predicted evolution of metallicity gradients from the TNG50 simulations \citep{Hemler2021}, the magenta line shows predictions from the FIRE-2 zoom-in simulations \citep{Sun2024} and the yellow line corresponds to the results from EAGLE simulations \citep{Tissera2022}. The grayscale shows a density histogram derived from previous observational results in the literature at $z= 0.5-3$ \citep{Swinbank2012, Queyrel2012, Jones2013, Leethochawalit2016, Wuyts2016, Molina2017, Carton2018, Schreiber2018, Patricio2019, Wang2017, Wang2019, Wang2020, Wang2022a, Curti2020,  Simons2021, Li2022, Venturi2024}. 
    The gradients in our sample are flat and nearly redshift-invariant, showing higher values than those from the MUGS, MAGICC, and TNG50 simulations, but similar to the results of the FIRE-2 zoom-in simulations.
    Right: Metallicity gradients as a function of stellar mass. The red points represent our results, and the red line represents the best linear fit. The blue and green lines correspond to local galaxies \citep[from][]{Belfiore2017} and star-forming galaxies at cosmic noon \citep[from][]{Cheng2024}, respectively. The magenta line is the mass-metallicity gradient relationship predicted by the FIRE-2 simulations. The grayscale density histogram in this panel is derived from results in the literature in the same range $z=0.5-1.7$ as our sample \citep{ Swinbank2012,Queyrel2012,Leethochawalit2016, Wuyts2016, Molina2017, Carton2018, Schreiber2018, Patricio2019, Wang2017, Wang2019, Wang2020, Wang2022a, Curti2020, Simons2021}. Our work reveals a mass dependence of metallicity gradients at $\sim$2-$\sigma$ significance, showing that more massive galaxies exhibit more negative gradients, consistent with the FIRE-2 simulation results.}
    \label{fig:grad}
\end{figure*}

According to the inside-out growth scenario, galaxies quickly form a compact core, resulting in very steep metallicity gradients in high-redshift galaxies. As galaxies evolve and their sizes increase, these gradients tend to flatten. This is similar to the predictions obtained with ``weak" feedback in the MUGS simulation \citep{Gibson2013}. In the same study, the MAGICC simulation with ``enhanced" feedback was presented. In this simulation, the stronger feedback distributes energy and recycles the interstellar medium (ISM) over larger scales, resulting in a ``flat" and nearly time-invariant metallicity gradient.
This demonstrates how the redshift evolution of gradient slopes is sensitive to feedback processes and galactic outflows.

The left panel of Figure~\ref{fig:grad} shows metallicity gradient slopes measured from our sample (red points) as a function of redshift, alongside other observational and theoretical results.
The orange and blue lines represent the MUGS and MAGICC simulations, respectively. These curves are for a single example galaxy. The green line corresponds to the results from the TNG50 simulation \citep{Hemler2021}, the yellow line illustrates EAGLE simulations \citep{Tissera2022}, and the magenta line shows the Feedback in Realistic Environments (FIRE-2) simulations \citep{Sun2024}. All these lines are for population averages. 
The FIRE-2 curve corresponds to a set of cosmological zoom-in simulations and shows the average metallicity gradient for eight galaxies with stellar masses ranging from ${\rm \log M_*/M_\odot} \approx 9$ -- 10.5 at $z=1$.
The grayscale is a density histogram of measurements from the observational literature at $z= 0.5-3$ \citep{Swinbank2012, Queyrel2012, Jones2013, Leethochawalit2016, Wuyts2016, Molina2017, Carton2018, Schreiber2018, Patricio2019, Wang2017, Wang2019, Wang2020, Wang2022a, Curti2020,  Simons2021, Li2022, Venturi2024}.
The metallicity gradients measured for our sample are relatively flat and nearly time-invariant over the observed redshift range, aligning with the results of other studies. They are slightly higher (more positive) than those predicted by the MUGS, MAGICC, and TNG50 simulations, but similar to the results of the FIRE-2 and EAGLE simulations. This suggests that feedback from star formation can significantly impact the metallicity gradient, particularly in galaxies experiencing intense star formation.

Tracing the growth of a galaxy through observations is nearly impossible. However, galaxies grow in mass as they evolve. 
Therefore, we can observe the changes in the metallicity gradient from small to high-mass galaxies as representing different stages in their growth \citep{Hemler2021}.
In the right panel of Figure~\ref{fig:grad}, we show the relationship between metallicity gradient slope and stellar mass (with masses listed in Table~\ref{table:1}). 
We find a significant correlation: more massive galaxies have more negative gradient slopes. We now quantify the mass dependence.
Considering that our sample consists of only 25 galaxies, we used bootstrap resampling to account for sample variance effects. Using 1000 iterations, we derived a linear model, resulting in the following equation:
\begin{equation}
\begin{split}
&\rm gradient (dex\ kpc^{-1})=\\
&\rm -0.0140 [\pm 0.0062] \times [log(M_*/M_\odot)-10]\\
&-0.0097 [\pm 0.0029].
   \label{eq:line}
\end{split}
\end{equation} 
The Pearson correlation coefficient is $-0.36$ and the P-value is $0.07$, indicating a fairly strong anti-correlation.
We calculated an RMSE of 0.01 dex~kpc$^{-1}$ for the slopes of these galaxies concerning this correlation. Our study therefore reveals a strong negative correlation between the metallicity gradient and galaxy mass, statistically significant at $\sim$2-$\sigma$.
More massive galaxies tend to be more extended, providing a greater number of usable spaxels for calculating the metallicity gradient. On average, 107 spaxels per galaxy are employed in this study, with more massive galaxies typically having more usable spaxels.
However, at fixed stellar mass, there is no clear correlation between the number of spaxels and the metallicity gradient slope. This supports the reliability of the relationship between stellar mass and the metallicity gradient (Equation~\ref{eq:line}). We consider the derived mass-metallicity gradient relation to be robust.

The root mean square (RMS) of the slopes for these 25 galaxies is 0.02 dex~kpc$^{-1}$, which is significantly smaller than the dispersion of the other observations in Figure~\ref{fig:grad} (grayscale). The overall RMS of these literature measurements is approximately 0.06 dex~kpc$^{-1}$, and we calculated the RMS of the slopes provided by various individual studies, resulting in a mean value of 0.05 dex~kpc$^{-1}$. Therefore, the dispersion of our sample is much smaller than that of the other observations in Figure~\ref{fig:grad} (grayscale), thanks to the high data quality enabled by the \msasd slit-stepping observing strategy.

The relationship between metallicity gradients and mass for local galaxies is illustrated with a blue line in Figure~\ref{fig:grad} (right panel). For low-mass galaxies, the gradient appears to be flat or slightly inverted. As the mass increases, the gradient steepens, and for massive galaxies above ${\rm \log M_*/M_\odot > 10.5}$ it becomes flatter again but remains negative \citep{Belfiore2017}. The flattening in massive galaxies may result from the metallicity reaching equilibrium, while the flattening in low-mass galaxies may be due to strong feedback, gas mixing, and wind recycling. At $z\sim1$, our sample shows that more massive galaxies exhibit more negative gradients, similar to the FIRE-2 simulation results. 
The flat gradients of the low-mass galaxies are similar to those found in local galaxies and can be attributed to strong stellar feedback.
However, we do not see a flattening of slopes in the most massive galaxies in our $z\sim1$ sample, and their steep slopes are not explained by the inside-out growth scenario. This difference at high masses compared to $z\sim0$ may be due to higher gas fractions and star formation rates at $z\sim1$, such that gas metallicities have not reached equilibrium and are more susceptible to processes such as radial gas flows.

Metallicity gradients are valuable indicators of the spatial distribution of metallicity within galaxies, but they do not capture all relevant characteristics (e.g., galaxy ID6199, ID9424 and ID11225). In galaxy ID9424, two distinct regions with varying metallicity are visible on the metallicity map, yet this is not reflected in the radial distribution alone. Azimuthal variations in metallicity, alongside radial gradients, may also provide critical insights. \cite{Bellardini2021, Bellardini2022} explored such azimuthal variations through FIRE simulations, showing that at early times (high redshift), local star formation leads to metal inhomogeneities in the azimuthal direction, and that the radial gradient may not fully describe the metallicity distribution. In galaxy ID9424, several strong H$\alpha$ clumps are located far from the galactic center and exhibit significantly lower metallicity. This may be attributed to gas accretion, which introduces azimuthal scatter without substantial radial variation. As galaxies evolve, they transition from being azimuthally dispersed to radially dominated systems, forming well-settled disks characterized by more pronounced radial distributions and greater azimuthal uniformity.
The non-radial variations in these \msasd metallicity maps can be used to infer ISM enrichment and mixing scales at $z\sim1$, and compare them with the $\sim1$ kpc scales found for $z=0$ \citep{Metha2021, Li2023}. We leave such an analysis for future work.

\subsection{The connection between galaxy chemo-structural evolution and disk formation: two case studies}

The evolution of metallicity gradients is thought to be closely related to the formation of well-established disks and thus the emergence of the Hubble sequence. Gas-phase gradients require limited radial mixing from processes such as mergers/interactions, inflows, and feedback-driven outflows, in addition to near-circular orbits (e.g., \citealt{Ma2017,Bellardini2021,Bellardini2022,Sun2024}). These conditions are also associated with disk settling.
In this section, we investigate the connection between metallicity gradients and galaxy disk dynamics with two case studies in the \msasd sample. In addition, Galaxy ID8512 is presented in \cite{Ivana2024}, showing a well-settled rotating disk together with a steep metallicity gradient. 
Galaxy ID9960 has the steepest gas-phase metallicity gradient in our sample (Table~\ref{table:1}), and exhibits a massive and regularly rotating disk. We also highlight the ability to measure its metallicity gradient despite the presence of an active galactic nucleus (AGN). 
In contrast, Galaxy ID4391 has a turbulent disk and a complex spatial metallicity distribution, possibly influenced by interaction with a nearby companion. 
These examples demonstrate our ability to probe the physical drivers of metallicity gradients. We plan to carry out a comprehensive study using our \msasd dataset in future work.

\subsubsection{Galaxy ID9960}

\begin{figure*}
    \centering
    \includegraphics[width=1.0\textwidth,clip,trim={0 0 0 0}]{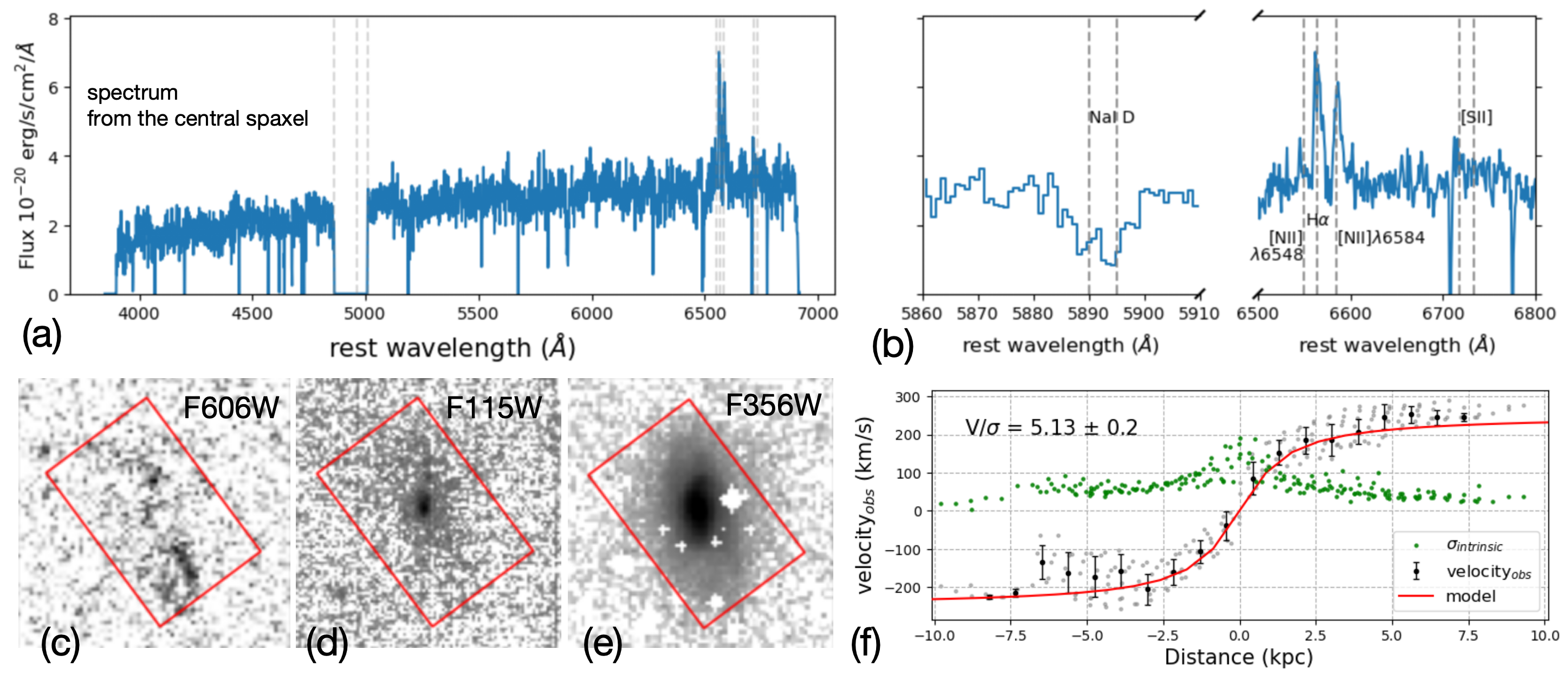}\\
    \caption{Detailed spectroscopic and kinematic properties of galaxy ID9960 --- the steep gradient source in our sample.
    Panel (a): The spectrum from the spaxel in the galaxy center, with dashed lines highlighting the emission lines of H$\beta$, \oiii, H$\alpha$, \nii, and \sii. H$\beta$ and \oiii\ unfortunately both fall within the chip gap.
    Panel (b): A zoomed-in spectrum of panel (a). The vertical dashed lines mark the wavelengths of Na\,{\scriptsize I} D$\lambda\lambda5890, 5896$, H$\alpha$, \nii, and \sii.
    The high flux ratio of \nii/H$\alpha$ and the strong blueshifted absorption of Na\,{\scriptsize I} D indicate the presence of AGN-driven outflows \citep[e.g.,][]{Dopita1998,Rupke2005, Davies2024}.
    Panels (c) to (e): image stamps of galaxy ID9960 in the filters of HST/ACS F606W, JWST/NIRCam F115W, and JWST/NIRCam F356W. The red box marks the entire field of view of our slit-stepping 3D spectroscopy. Panel (f): The one-dimensional velocity profile (black and grey dots) with the best-fitting model shown in red. The green dots represent the intrinsic velocity dispersion, corrected for the instrument line spread function as described in the text. 
   We derive $\rm v_{c}/\sigma_0 = 5.13\pm 0.20$, which together with the two-dimensional velocity map (Figure~\ref{fig:1}) indicates that galaxy ID9960 has a regularly rotating disk. 
    The steep metallicity gradient and the clear rotation signatures of galaxy ID9960 supports the strong connection between chemo-structural evolution and disk formation.}
    
    \label{fig:9960}
\end{figure*}

Among the 25 galaxies examined in this work, galaxy ID9960 at $z=1.51$ exhibits a very steep negative metallicity gradient, surpassing even that of local galaxies (see Figs.~\ref{fig:grad} and \ref{fig:1}). It has a stellar mass of log(M$_{\text{*}}$/M${_\odot}$) = 11.20, the most massive galaxy within our sample. 
In panel (a) of Figure~\ref{fig:9960}, we present the spectrum from the central spaxel of this galaxy.
Unfortunately, the H$\beta$ and \oiii\ lines fall in the chip gap, such that we cannot determine the star formation and AGN regions through the BPT diagram. 
We instead use the WHAN diagnostic diagram. As shown in Figure~\ref{fig:1}, the central region of this galaxy is classified as a sAGN (i.e., Seyfert), while the outer regions are classified as star forming. We conclude that there is a complex excitation structure comprising both star formation and a central AGN, which can be separated thanks to the angular resolution and minimal beam smearing of these data. 
We reiterate that the derived metallicity gradient slope is based exclusively on the spaxels classified as star forming, where the adopted strong line calibrations are applicable.

To further examine the nature of the nuclear activity, we plot the spectrum of the central spaxel zoomed in around H$\alpha$, \nii, and \sii\ in panel (b) of Figure~\ref{fig:9960}. The spectrum exhibits narrow line emission, with a velocity dispersion of about 160 km~s$^{-1}$ for H$\alpha$ at the galaxy center. 
We do not observe a strong broad ($\sigma \gtrsim 1000$~km~s$^{-1}$) emission component, as is seen in Type 1 AGNs. The spectral features thus indicate that this galaxy hosts a Type 2 AGN.
Furthermore, panels (c) to (e) of Figure~\ref{fig:9960} present images obtained from the HST/ACS F606W (rest-frame wavelength $\sim$2400~\AA), JWST/NIRCam F115W\footnote{https://dawn-cph.github.io/dja/index.html} (rest-frame $\sim$4600~\AA), and JWST/NIRCam F356W (rest-frame $\sim$1.41~$\mu$m) filters \citep{Koekemoer2011, Wang2023}. While the central region remains indistinct in the rest-frame UV image (F606W), a faint but discernible disk structure is apparent. The lack of a UV-bright nucleus is also indicative of a Type 2 AGN.

We now consider the gas kinematics of galaxy ID9960, to examine whether the strong metallicity gradient is associated with a rotation-dominated star forming disk. While we see evidence of AGN emission and possible AGN-driven outflows (Figure~\ref{fig:9960}), these signatures are confined to the central region. The extended emission is dominated by star formation.
As shown in Figure~\ref{fig:1}, ID9960's H$\alpha$ velocity map clearly exhibits a mature rotating disk. It satisfies all standard disk classification criteria \citep[e.g.,][]{schreiberSINSZCSINFSurvey2018} including a ``spider diagram'' pattern in the velocity field. The rotation curve extracted along the major axis is shown in panel (f) of Figure~\ref{fig:9960}. 
To correct the observed (line-of-sight) velocity map for inclination angle $i$, we use the axis ratio $b/a$ assuming a thin disk: 
$\cos(i) = b/a$. The axis ratio of 0.48 from the CANDELS photometric catalog leads to $i=61^\circ$. 
We model the rotation following the method described in \cite{Ju2022}, adopting an arctangent rotation curve \citep[e.g.,][]{Jones2010}:
\begin{equation}
	V(R) = v_0 + \frac{2}{\pi} v_{c} \arctan(R/R_{t}).
	\label{eq:vel}
\end{equation}
Here $R_{t}$ is the scale radius, $v_{c}$ is the asymptotic rotation velocity and $v_{0}$ is an overall systemic velocity.
We use the nested sampling code \texttt{nautilus} \citep{Lange2023} to find the best-fit rotating disk model, presented in panel (f) of Figure~\ref{fig:9960}. Black dots represent the median observed velocity within the distance bins, while the red line corresponds to the best-fit model. 
The model matches the data to within a RMSE of 33 km~s$^{-1}$, comparable to the measurement scatter. 
The $v_c = 250_{-5}^{+6}$~km~s$^{-1}$ is similar to the Milky Way ($\sim 230$~km~s$^{-1}$; e.g., \citealt{Bland-Hawthorn2016}). The estimated dynamical mass within the half-light radius (6 kpc) is approximately $10^{11} \rm{M_\odot}$, compatible with the stellar mass and a modest fraction of gas and dark matter.

The ratio of the rotation velocity of a gas disk to its velocity dispersion, $v_c/\sigma$, is a key indicator of dynamical support. Typically, the calculation uses $ v_c$ as the maximum rotational velocity and $\sigma_0$ as the intrinsic velocity dispersion from the outer regions of galaxies \citep[where rotation curves are flat; e.g.,][]{Wisnioski2015}. In this work, we calculate the ratio of $ v_{c}$ to the velocity dispersion $\sigma_0 = 48.9 \pm 1.6$~km~s$^{-1}$ (median and sample standard error) of the spaxels located at radius $> 3$~kpc.
The resulting ratio is $v_{c}/\sigma_0 = 5.13 \pm\ 0.20$, which is typical of massive $z\sim1$ galaxies. 
According to the classifications of \cite{Girard2020} and \cite{Kassin2012}, for example, the $ v/\sigma$ ratio is used to classify galaxies as ``rotationally supported" when $ v/\sigma > 1$. Regular or well-settled rotation is identified by $ v/\sigma > 3$, while $ 1 < v/\sigma < 3$ indicates irregular or disturbed rotation. Hence, the $ v/\sigma$ of this galaxy is high enough to qualify as regularly rotating. 

In summary, galaxy ID9960 exhibits a steep metallicity gradient and a well-settled star-forming disk. This supports a physical picture in which strong rotational support is necessary for sustaining a radial abundance gradient. This galaxy represents a compelling case at $z\approx1.5$ for co-evolution of chemical profiles and gas kinematics, signifying the emergence of a modern Hubble sequence.

\subsubsection{Galaxy ID4391}

\begin{figure*}
    \centering
    \includegraphics[width=1.0\textwidth,clip,trim={0 0 0 0}]{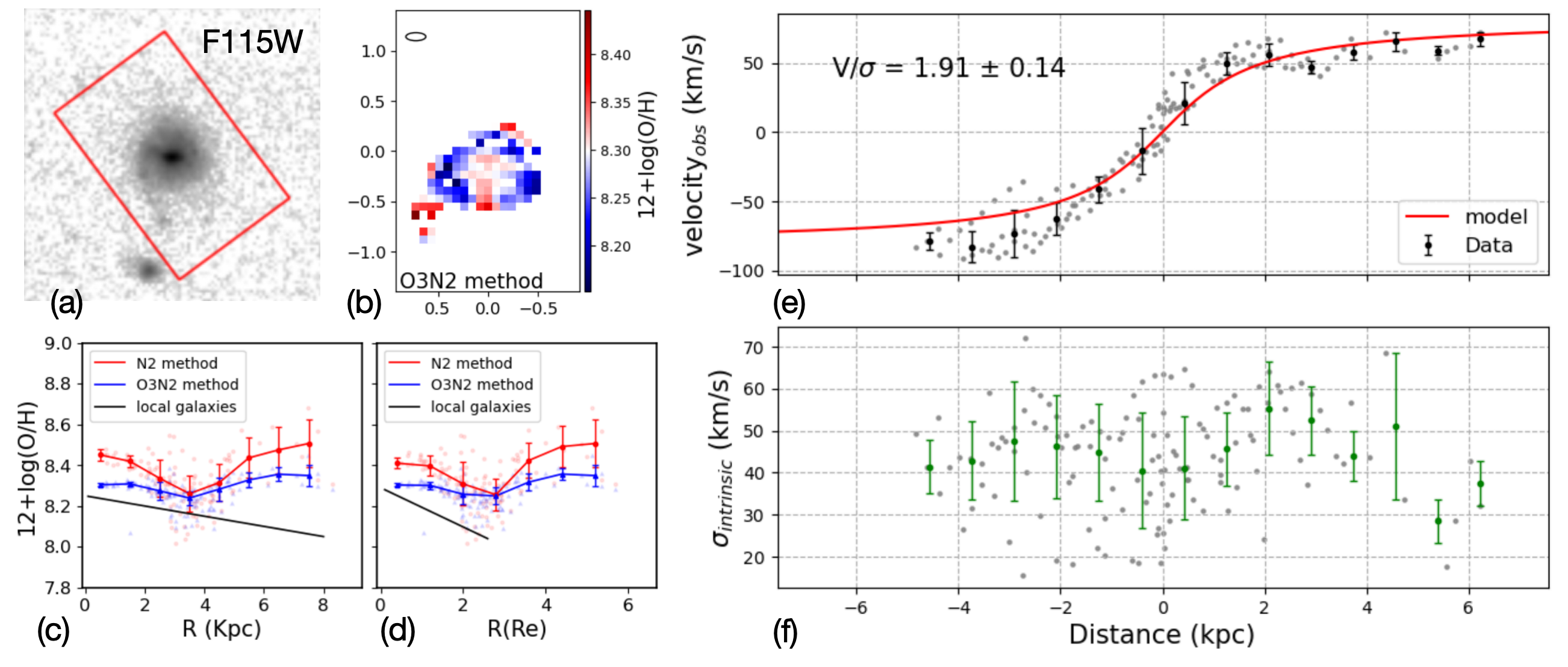}\\
    \caption{Kinematic properties of galaxy ID4391, with complex chemical profile verified by different metallicity indicators.
    Panel (a): the NIRCam/F115W image stamp of galaxy ID4391. Panel (b): the metallicity map obtained using the O3N2 method. Panels (c) and (d): metallicity gradients with galactocentric radii in units of kpc and $R_e$ units determined by the N2 method (red dots) and the O3N2 method (blue dots). The points and error bars show the metallicity profiles using averages in radial bins. The black line in the two panels represents the slope for local galaxies \citep{Belfiore2017}. Both methods display a distinct gradient from the galaxy center to the outskirts, with a turning point located near 2.5 $R_e$. Panel (e): The one-dimensional velocity profile (black and grey dots) with the best-fitting model is shown in red. 
    Panel (f): The one-dimensional intrinsic velocity dispersion is presented on the same spatial axis as the velocity profile. The gray dots represent the intrinsic velocity dispersion, while the green error bars indicate the median and standard deviation of the gray dots within radial bins. The $\rm v_{c}/\sigma$ is $1.91 \pm\ 0.14$, indicating irregular rotational motion and suggesting that this galaxy likely has a thick disk. The unusual metallicity profile might be attributed to galaxy mergers or gas inflows, both of which are also capable of significantly disturbing gas motions.}
    
    \label{fig:4391}
\end{figure*}

In contrast to galaxy ID9960 introduced above, galaxy ID4391 shows a more complex chemical profile and more turbulent gas kinematics, possibly affected by interaction with a neighboring galaxy.
First of all, its metallicity ${\rm \log(O/H)}$ does not follow a linear trend with radius (Figure~\ref{fig:4391}), such that the derived metallicity gradient depends on the radial range used.
Within the range 0.5 -- 3.5 kpc ($\rm \sim 2.5~R_e$), the data are well described with a gradient of best-fit slope $\rm -0.074 \pm 0.012\ dex\ kpc^{-1}$.
However, the metallicity increases at $R \gtrsim 5$~kpc. Fitting the range 0.5 -- 10 kpc, the gradient ``flattens'' to $\rm 0.0013 \pm 0.0117\ dex\ kpc^{-1}$.

The quality of our \msasd IFS data enables a detailed characterization of the complex metallicity spatial distribution. As shown in the 2D metallicity map in panel (b) of Figure~\ref{fig:4391}, the increasing O/H at large radius is largely driven by a tail-like structure extending to the lower left, with metallicity comparable to that found in the galaxy's center. We verify that this structure is robust by additionally measuring metallicity using the O3N2 ratio (as calibrated by \citealt{marino2014}). Panels (c) and (d) of Figure~\ref{fig:4391} show the same general trend with both the N2 (red) and O3N2 (blue) diagnostics. 
This tail structure, spanning a galactocentric radius ${\rm R \simeq 2.5 - 5~R_e}$, is connected to a neighboring galaxy visible in the image cutout. While this source falls outside our IFS coverage (see panel (a) of Figure~\ref{fig:4391}) and is not spectroscopically confirmed, its photometric redshift $z\sim1$ is similar to ID4391, with a mass ratio of approximately 0.2 \citep{Yung2022}. We speculate that the high-metallicity tail may be related to a merger or gas flows associated with this neighboring galaxy.

To explore the dynamical evidence of possible merger-induced turbulence, we examine the velocity and velocity dispersion maps of galaxy ID4391 shown in Figure~\ref{fig:1}. The velocity dispersion is clearly amplified along the direction to the possible merging companion, which roughly corresponds to the kinematic major axis. Using the same method as for ID9960 above, we constructed a rotating disk model for this galaxy. 
The velocity and dispersion profiles are shown in panels (e) and (f) of Figure~\ref{fig:4391}, along with the best-fit disk model. We obtain $ v_c = 82_{-3}^{+4}$~km~s$^{-1}$ with an inclination angle $ i = 67_{-4}^{+3}$ degrees. The median intrinsic velocity dispersion $\sigma_0$ of the spaxels located beyond 3 kpc (where the rotation curve is approximately flat) is $ 42.7\pm1.8$ km~s$^{-1}$.
The $ v_{c}/\sigma_0$ ratio is $1.91 \pm 0.14$, indicating irregular rotational motion according to classification criteria, consistent with a thick gas disk, likely perturbed by gravitational interactions from the companion galaxy.

\section{Summary}\label{summary}

In this paper, we analyze 3D spectroscopy of 43 star-forming galaxies at redshifts $z=0.5$--1.7 obtained using a novel multiplexed slit-stepping approach via the \msasd survey \citep{Ivana2024}.
The observations cover multiple strong rest-frame optical emission lines (e.g., H$\beta$, \oiii, H$\alpha$, \nii, and \sii) with spectral resolution $R\sim2700$. 
This allows us to spatially map the emission line flux, EWs, velocity field, and velocity dispersion. From the line flux ratios, we investigate the gas-phase metallicity distributions and analyze trends in the radial metallicity gradients.

We focus on the subset of 25 galaxies in our sample with homogeneous metallicity gradients measured using the N2 method. The gradient slopes range from $-$0.03
to 0.02 dex~kpc$^{-1}$, with most galaxies showing negative or flat gradients. The gradients show a clear negative correlation with stellar mass, statistically significant at the $\sim 2\sigma$ level, with more massive galaxies having steeper gradients. The scatter of the gradient slope measurements is very small, with an RMS of 0.02 dex~kpc$^{-1}$. Compared with simulation results, we find that the gradients are similar to predictions from the FIRE-2 simulations, which incorporate feedback in cosmological galaxy formation.
Relatively flat gradients in low-mass galaxies can be attributed to strong, time-dependent stellar feedback. 
The steeper gradients in more massive galaxies in our sample are not explained by a simple inside-out growth scenario. 
The steep gradients may arise from high gas fractions and star formation rates at $z\sim1$ compared to $z\sim0$, which prevent the metallicity from reaching equilibrium and make it more susceptible to effects such as radial gas flows.

We highlight two exemplary galaxies within our sample, illustrating the ability of these data to perform both sample-wide analyses and detailed case studies.
Galaxy ID9960 has the most negative metallicity gradient and highest stellar mass within the sample, with log(M$_{\text{*}}$/M$_{\odot}$) = 11.20. It contains a rotation-dominated star forming disk with $\rm v_{c}/\sigma = 5.13\pm0.20$.
The chemical profiles and gas kinematics of this massive $z\sim1.5$ disk galaxy likely signify the end stage of the co-evolution of chemical profiles and disk structure, where the steep abundance gradient is maintained by strong rotational support.
In contrast, galaxy ID4391 shows a complex spatial metallicity distribution which is not adequately described by a simple radial gradient. 
The gradient is steeply negative within 2.5 effective radii, while beyond this range it becomes positive. This galaxy features a thick rotating disk, with increased velocity dispersion in the direction of a possible interacting companion. 
Its gradient might be affected by gas accretion induced by merging. 
These galaxies may offer key insight into the tight connection between galaxy chemo-structural evolution and disk assembly, as the modern Hubble sequence emerges.

This paper presents demographic trends of metallicity gradients in the $z\sim1$ galaxy population. This work takes advantage of MSA multiplexing to secure IFS datasets for a sample of 43 galaxies simultaneously, from which we obtain 25 robust metallicity gradient measurements. We find a tight relationship between stellar mass and metallicity gradient at $z\sim1$, which provides a crucial benchmark for advancing our understanding of galaxies' chemical evolution.
Notably, the high-quality data provided by \msasd were obtained with only $\sim$ 30 hours of JWST time. An equivalent survey with NIRSpec's IFU would require an order of magnitude more time. This work thus demonstrates the remarkable efficiency gain of our slit-stepping strategy (see also \citealt{Ivana2024}). To date, three other JWST programs have adopted a similar slit-stepping approach (i.e., GO-3426 (PI: Jones), GO-2123 (PI: Kassin) and GO-4291 (PI: Kassin)). We envision further programs to expand the available sample in terms of galaxy demographics and redshift, enabling a more comprehensive study of the correlation between the chemical evolution of galaxies and the properties of their disks across their most formative epochs.

\section*{Acknowledgements}
We thank the anonymous referee for the constructive comments, which significantly improved the manuscript.
This work is supported by the National Natural Science Foundation of China (grant 12373009), the CAS Project for Young Scientists in Basic Research Grant No. YSBR-062, the Fundamental Research Funds for the Central Universities, and the science research grant from the China Manned Space Project. XW acknowledges the support by the Xiaomi Young Talents Program, and the work carried out, in part, at the Swinburne University of Technology, sponsored by the ACAMAR visiting fellowship.
This work is based on observations made with the NASA/ESA/CSA James Webb Space Telescope. The data were obtained from the Mikulski Archive for Space Telescopes at the Space Telescope Science Institute, which is operated by the Association of Universities for Research in Astronomy, Inc., under NASA contract NAS 5-03127 for JWST. These observations are associated with program JWST-GO-2136. We acknowledge financial support from NASA through grant JWST-GO-2136. 
TJ acknowledges support from a Chancellor's Fellowship and a Dean's Faculty Fellowship, and from NASA through grant 80NSSC23K1132. 
CAFG was supported by NSF through grants AST-2108230 and AST-2307327; by NASA through grant 21-ATP21-0036; and by STScI through grant JWST-AR-03252.001-A.

\facilities{JWST (NIRSpec MSA)}
\software{nautilus \citep{Lange2023}}

The JWST data presented in this article were obtained from the Mikulski Archive for Space Telescopes (MAST) at the Space Telescope Science Institute. The specific observations analyzed can be accessed via \dataset[doi:10.17909/s8wp-5w10]{https://doi.org/10.17909/s8wp-5w10}.

\appendix
\section{The spatially resolved 2D maps of our sample galaxies}
\label{sec:appendix}
\restartappendixnumbering
\begin{figure*}
    \centering
    \includegraphics[width=0.8\textwidth,clip,trim={0 0 0 0}]{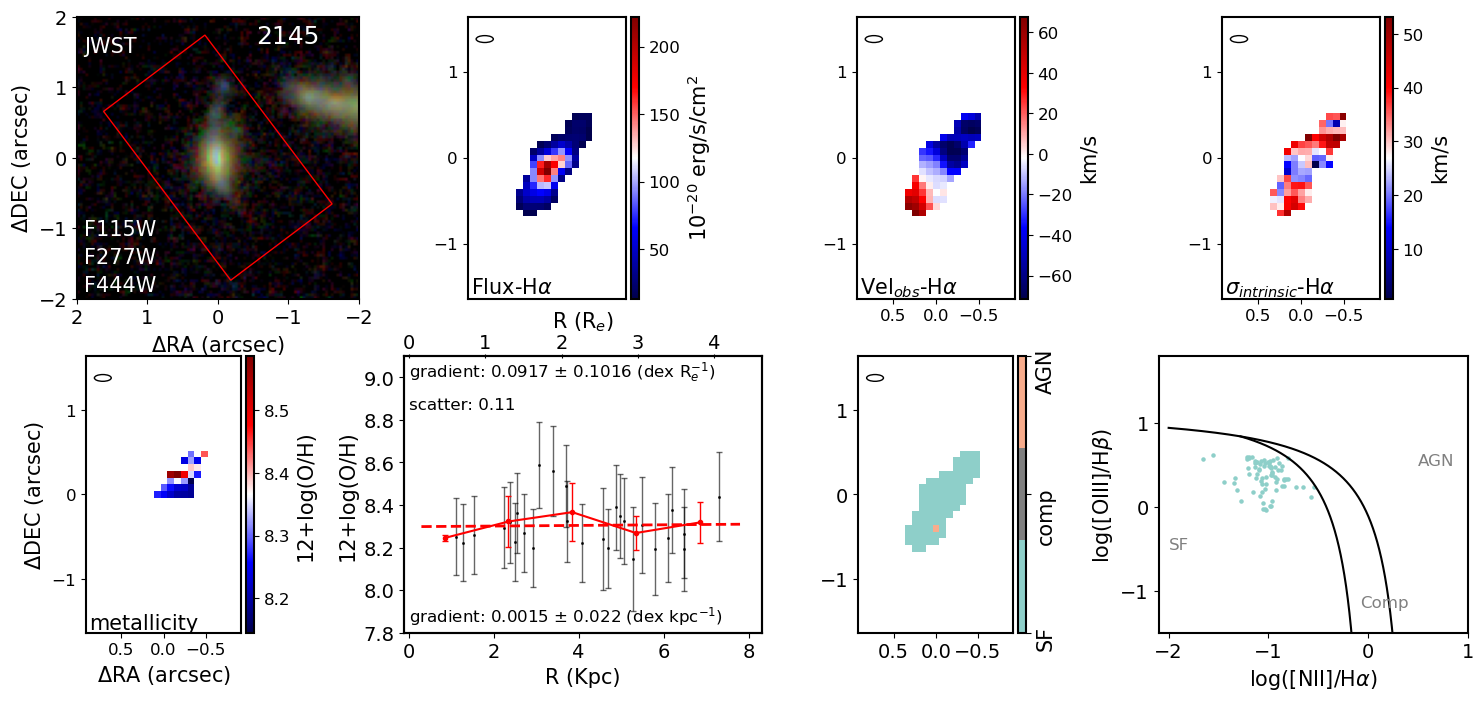}\\
    \includegraphics[width=0.8\textwidth,clip,trim={0 0 0 0}]{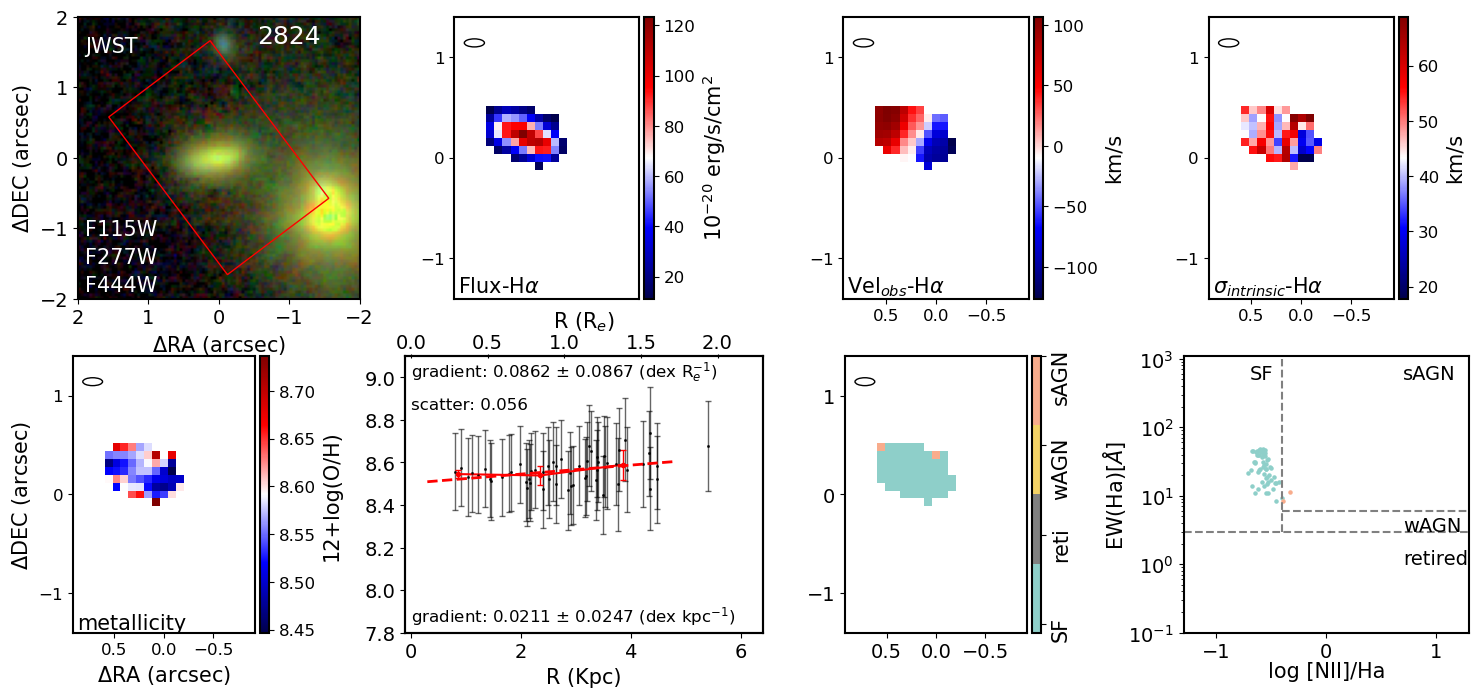}\\
    \includegraphics[width=0.8\textwidth,clip,trim={0 0 0 0}]{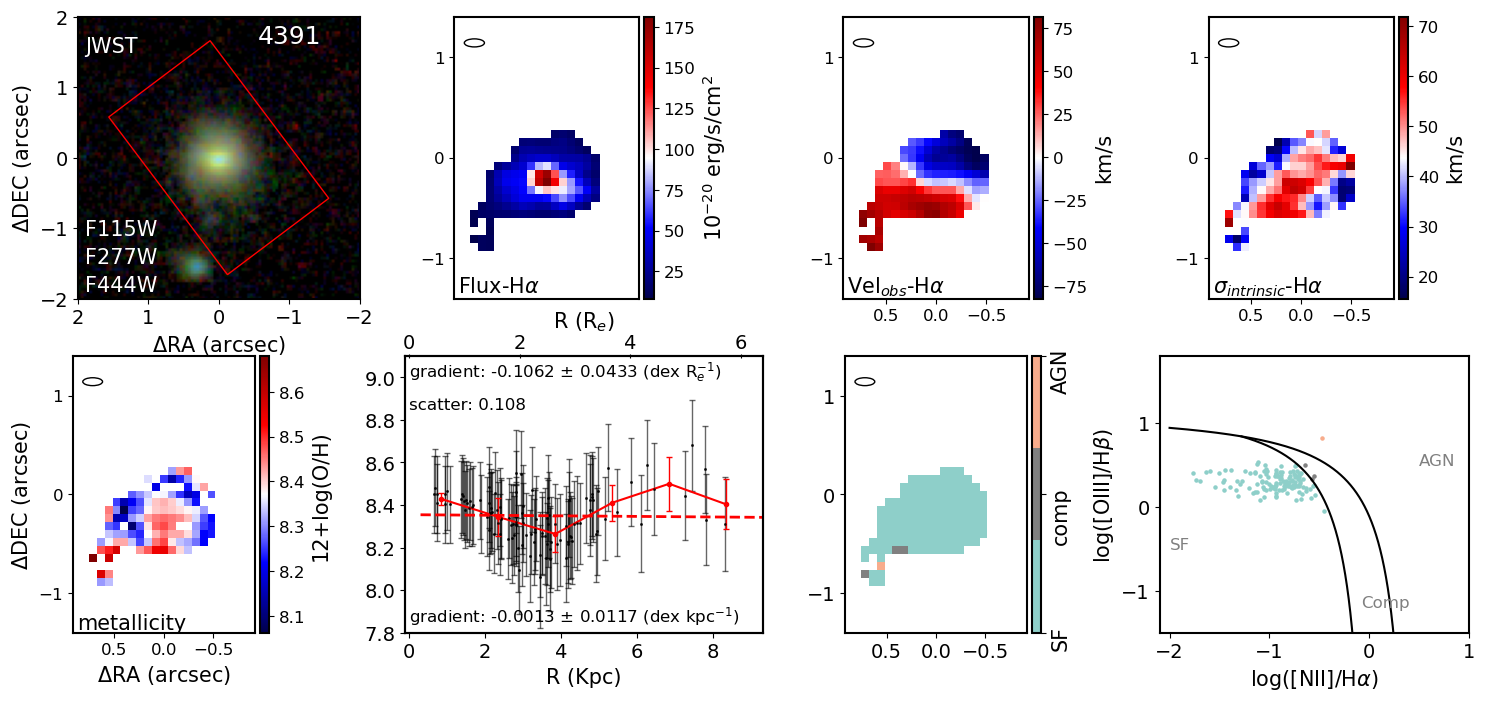}\\
    \caption{A comprehensive spatially-resolved view of our galaxy sample with metallicity gradient measurements. For each galaxy we show 8 subplots on two rows. 
    The first row displays, from left to right:
    the 3-color image, H$\alpha$ emission flux map, observed H$\alpha$ velocity map, and intrinsic H$\alpha$ velocity dispersion map (corrected for instrument resolution). The second row shows the gas-phase metallicity map using the N2 method, the metallicity gradient with the best linear fit represented by the red dashed line, a map color-coded according to the positions of individual spaxels in the BPT or WHAN diagram, and the BPT or WHAN diagram used to classify the ionization properties in each galaxy. 
    The red boxes in the three-color images indicate the FoV ($1\farcs8\times(2\farcs0-3\farcs0)$) for each of our sources. The ellipses on the maps represent the size of the resolution element of our \msasd\ integral field spectroscopy.  
}
    \label{fig:1}

\end{figure*}

\begin{figure*}
\ContinuedFloat
\centering
    \includegraphics[width=0.8\textwidth,clip,trim={0 0 0 0}]{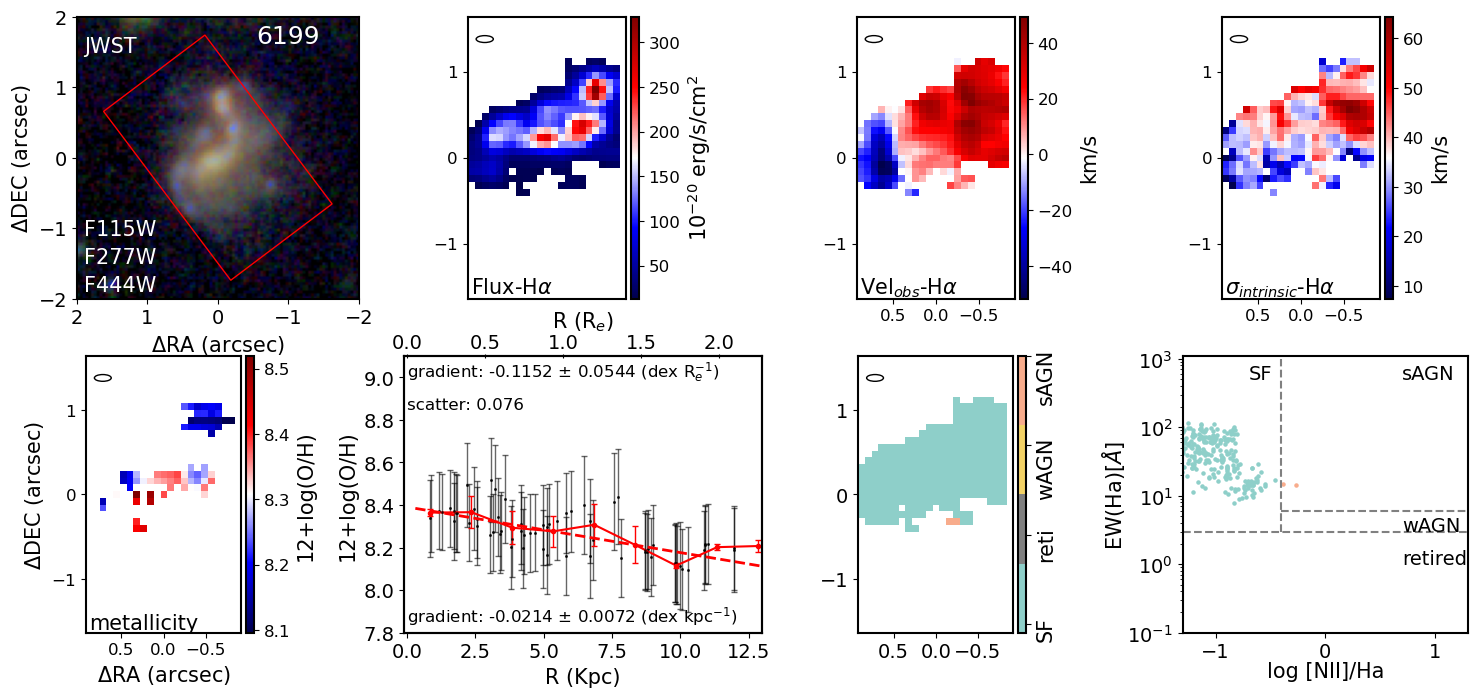}\\
    \includegraphics[width=0.8\textwidth,clip,trim={0 0 0 0}]{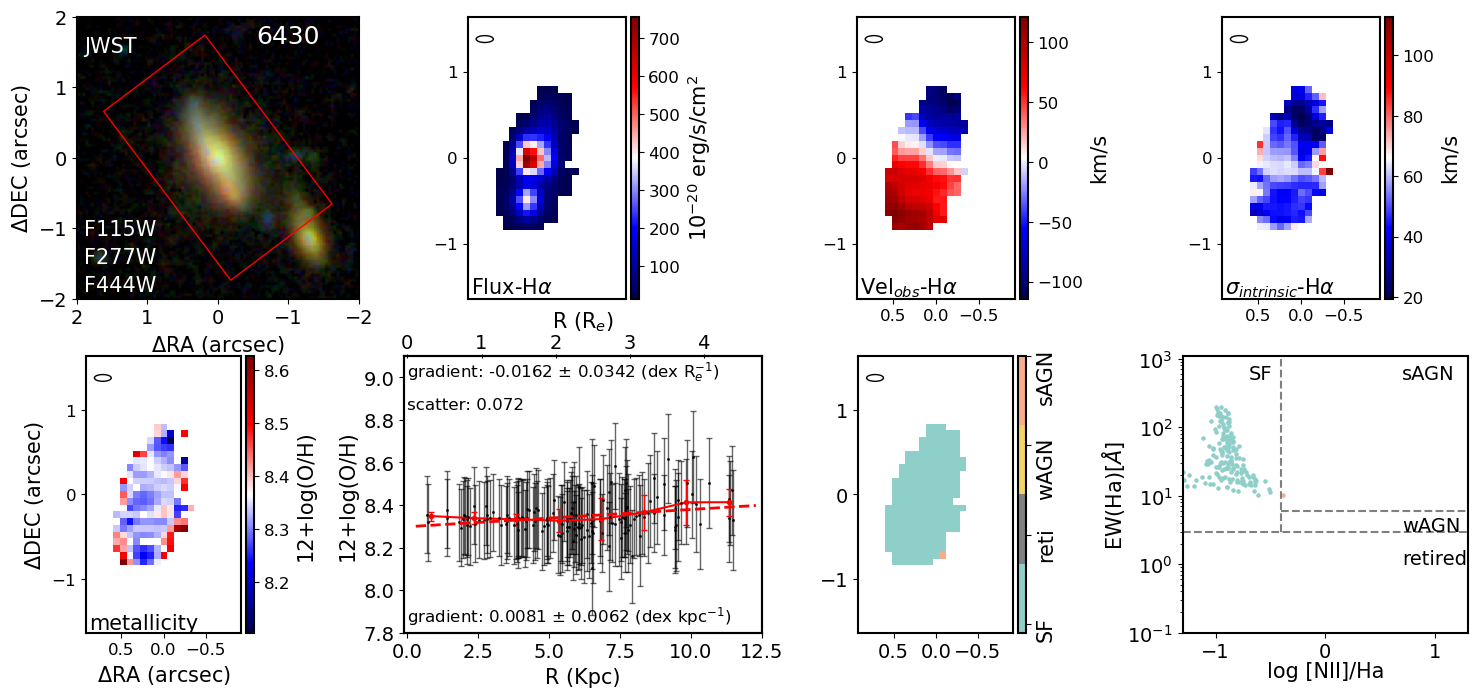}\\
    \includegraphics[width=0.8\textwidth,clip,trim={0 0 0 0}]{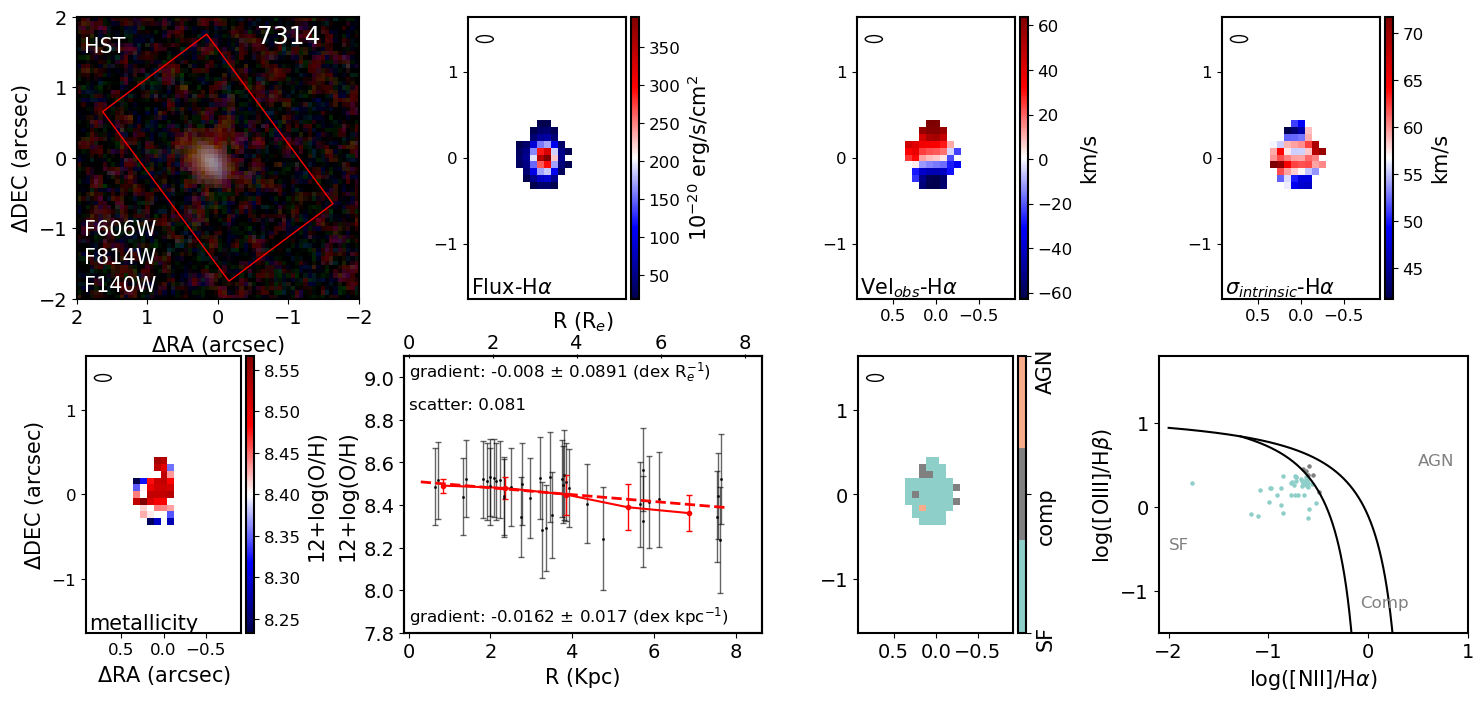}\\
    \caption{
    (continued) For subplots other than the metallicity maps and the metallicity gradients, we show spaxels with $\rm S/N\ of\ Flux(H\alpha) > 10$ and $\rm S/N\ of\ EW(H\alpha) > 5$. For the metallicity maps and gradients, we plot only the subset of those spaxels classified as star-forming with statistical uncertainty of $< 0.25$ dex in metallicity. 
    For the metallicity gradient subplots, the bottom and top abscissas are the galactocentric radius in units of kpc and effective radius $\mathrm{R_e}$, respectively.
    In these subplots we report the best-fit metallicity gradient slope for each galaxy based on individual spaxels, along with the scatter (i.e., the RMSE between metallicity of individual spaxels and the linear fit).
    The red points and error bars show the metallicity profiles using averages in radial bins.}
\end{figure*}
\begin{figure*}
\centering
    \ContinuedFloat
    \includegraphics[width=0.8\textwidth,clip,trim={0 0 0 0}]{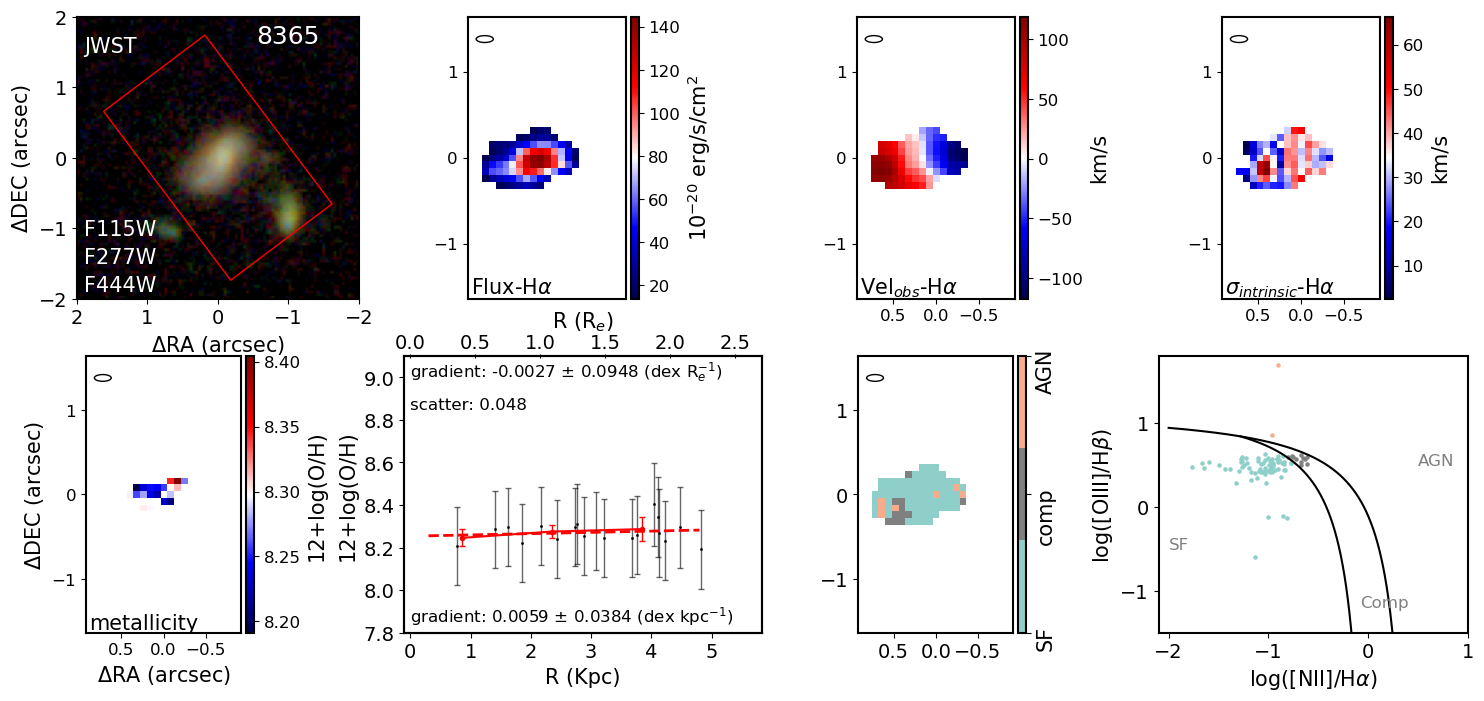}\\
    \includegraphics[width=0.8\textwidth,clip,trim={0 0 0 0}]{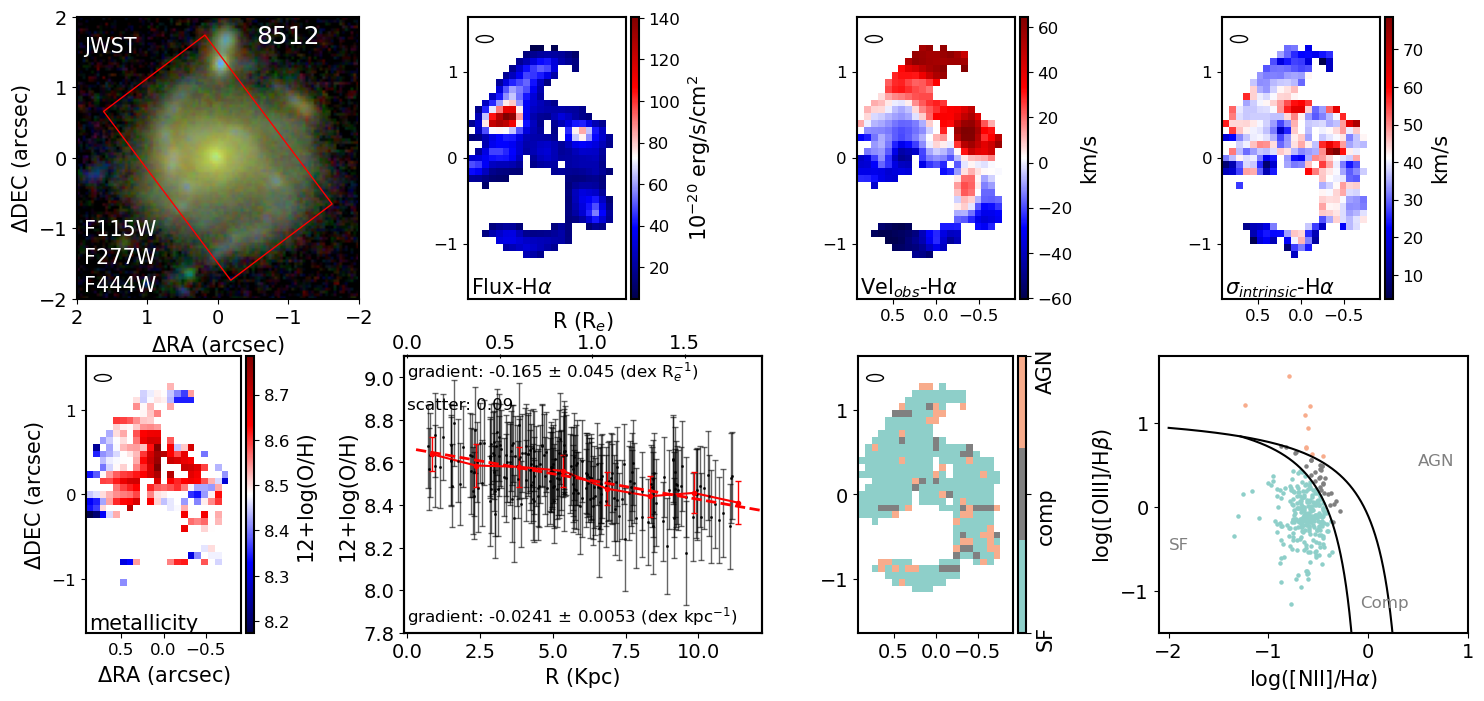}\\
    \includegraphics[width=0.8\textwidth,clip,trim={0 0 0 0}]{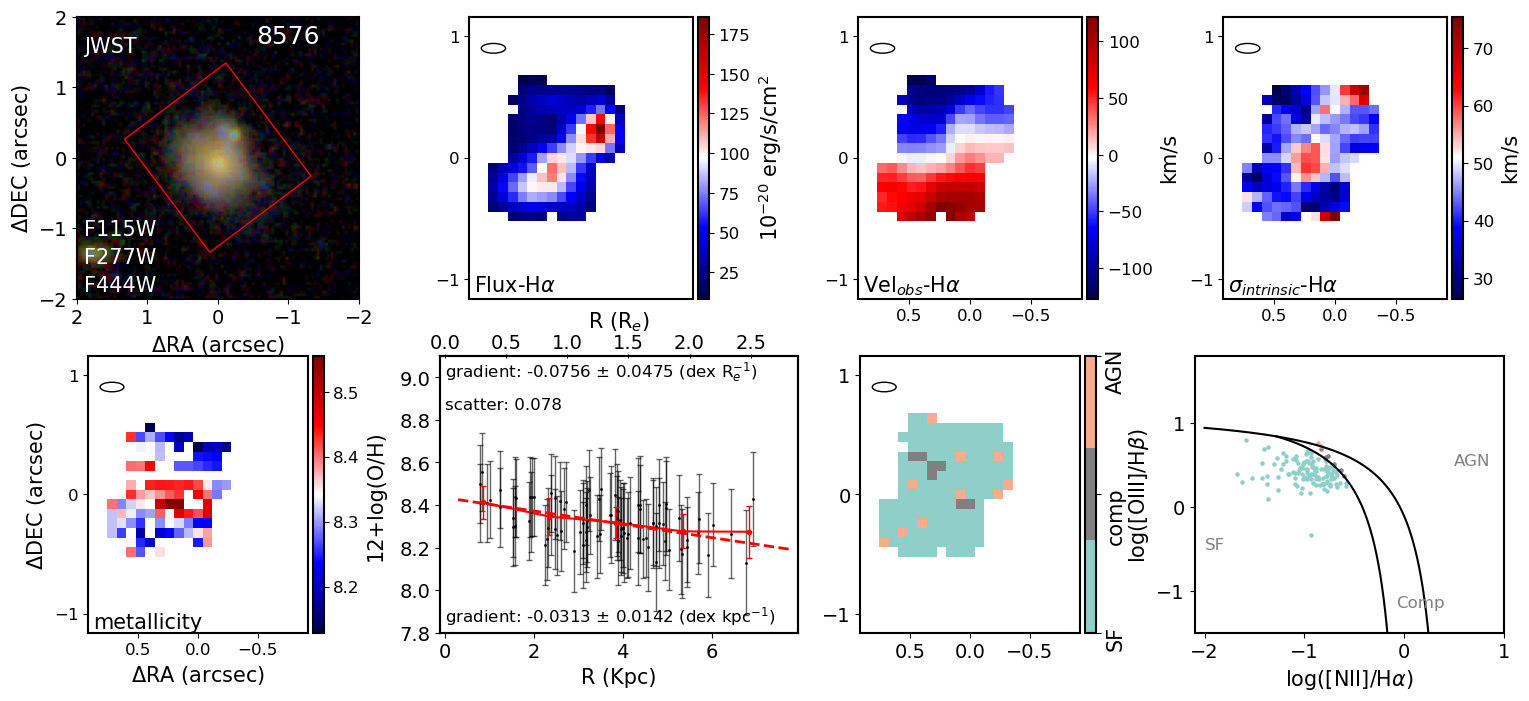}\\
    {\bf Figure A1.} continued\\
\end{figure*}
\begin{figure*}
\centering
    \ContinuedFloat

    \includegraphics[width=0.8\textwidth,clip,trim={0 0 0 0}]{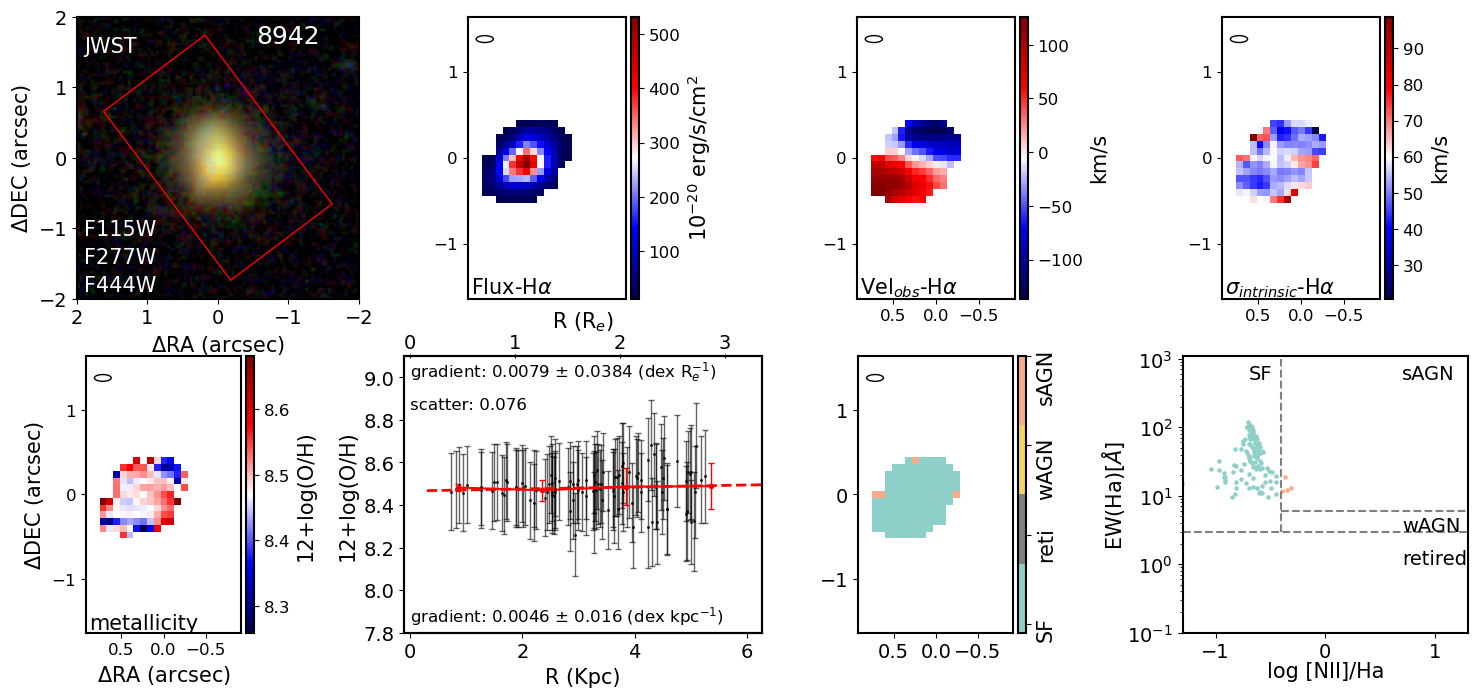}\\
    \includegraphics[width=0.8\textwidth,clip,trim={0 0 0 0}]{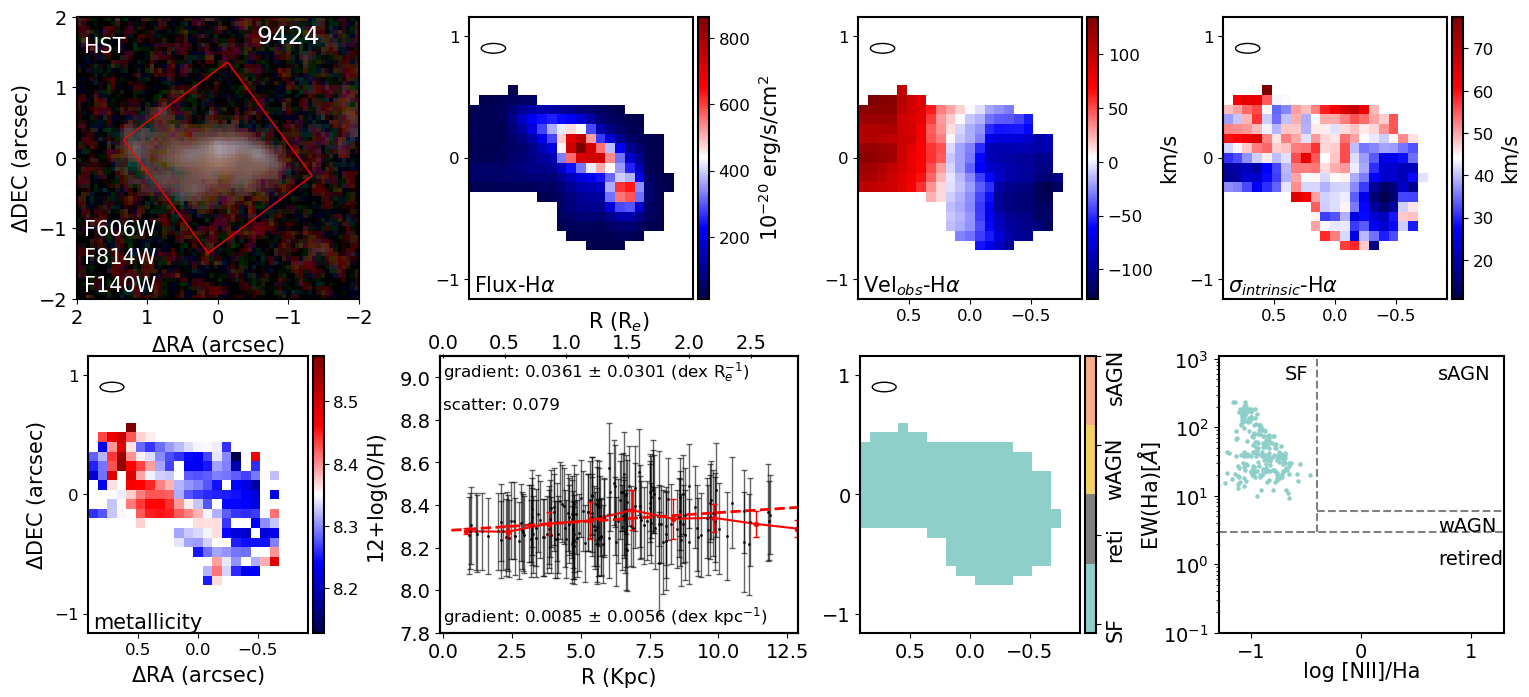}\\
    \includegraphics[width=0.8\textwidth,clip,trim={0 0 0 0}]{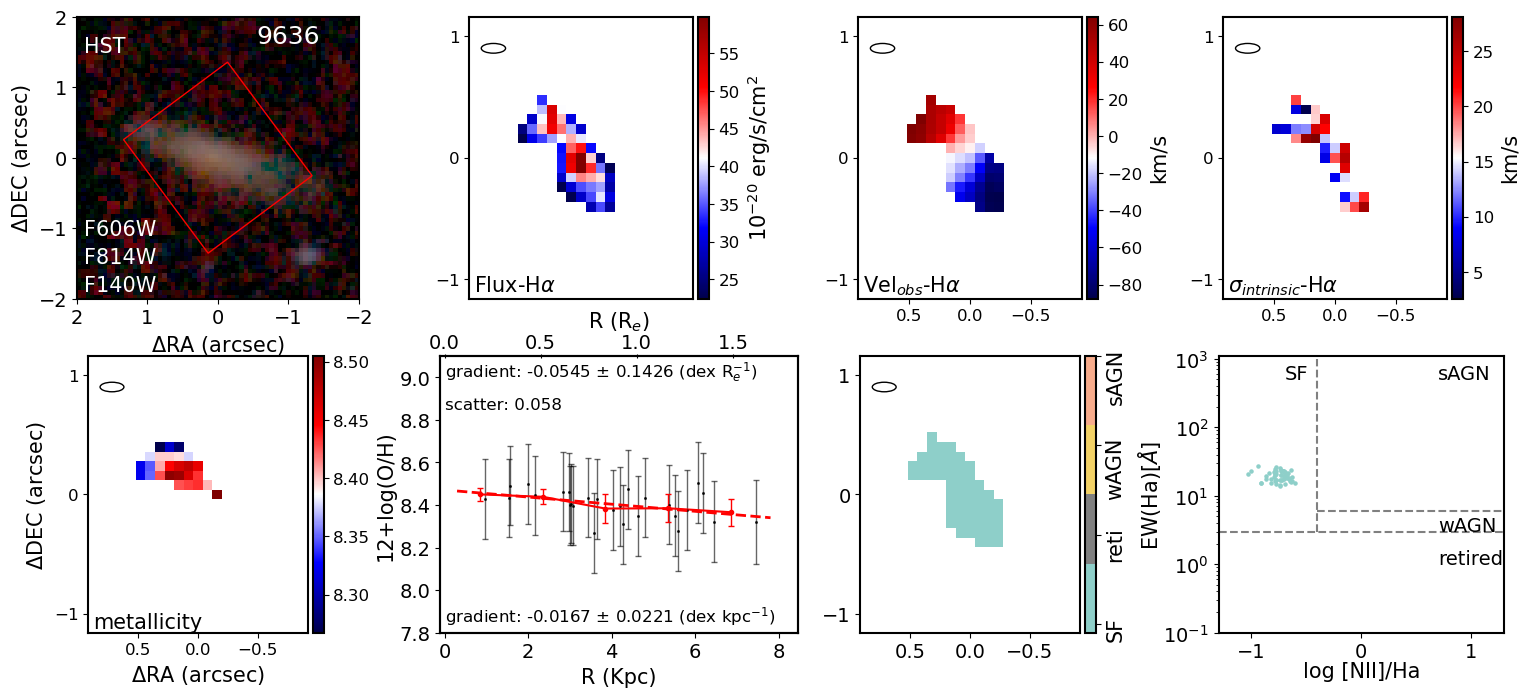}\\
    {\bf Figure A1.} continued\\
\end{figure*}
\begin{figure*}
\centering
    \ContinuedFloat

    \includegraphics[width=0.8\textwidth,clip,trim={0 0 0 0}]{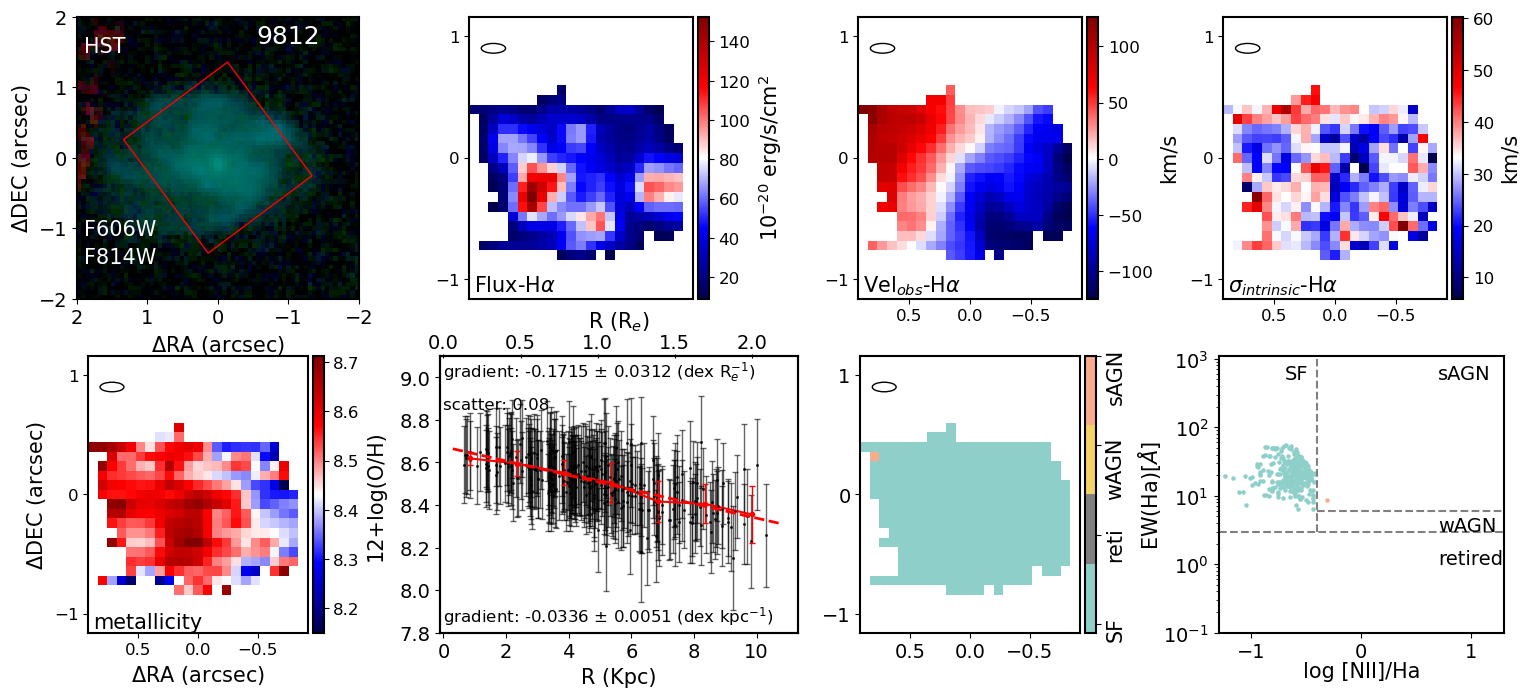}\\
    \includegraphics[width=0.8\textwidth,clip,trim={0 0 0 0}]{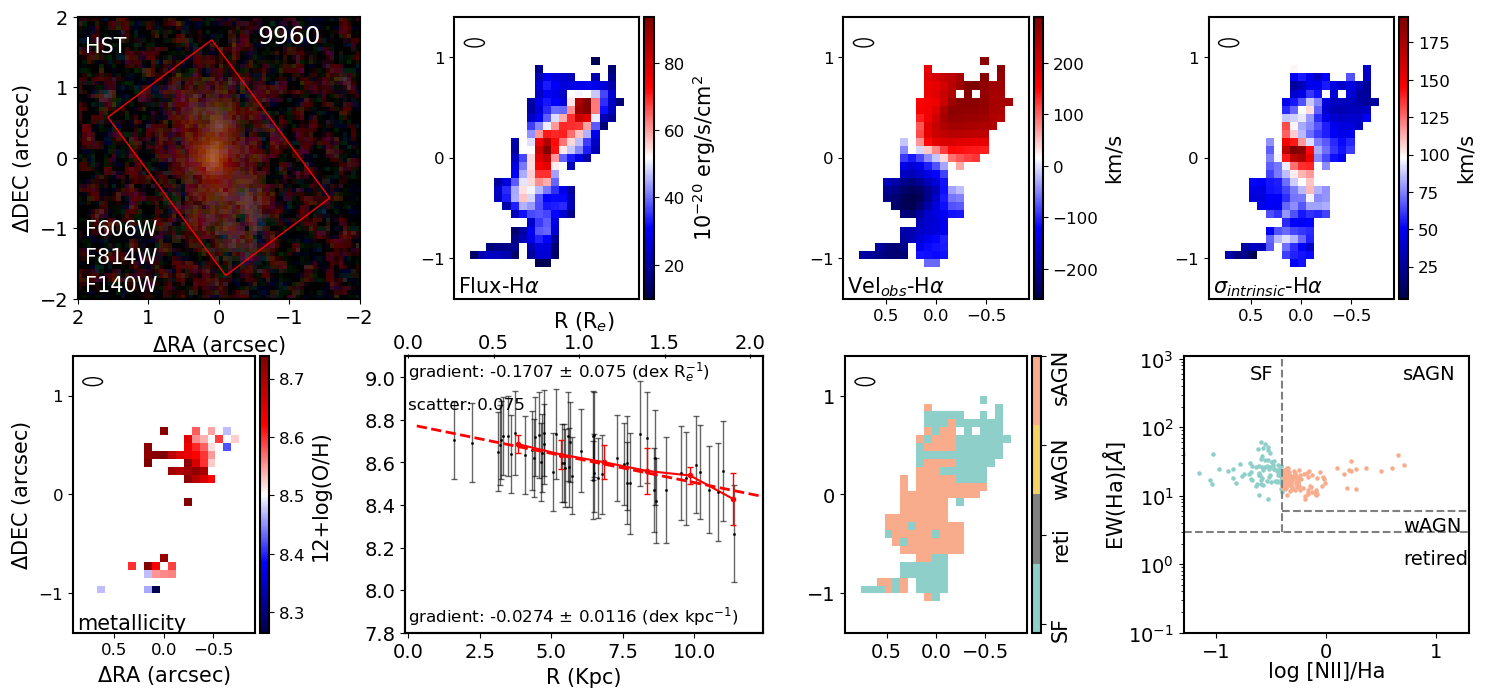}\\
    \includegraphics[width=0.8\textwidth,clip,trim={0 0 0 0}]{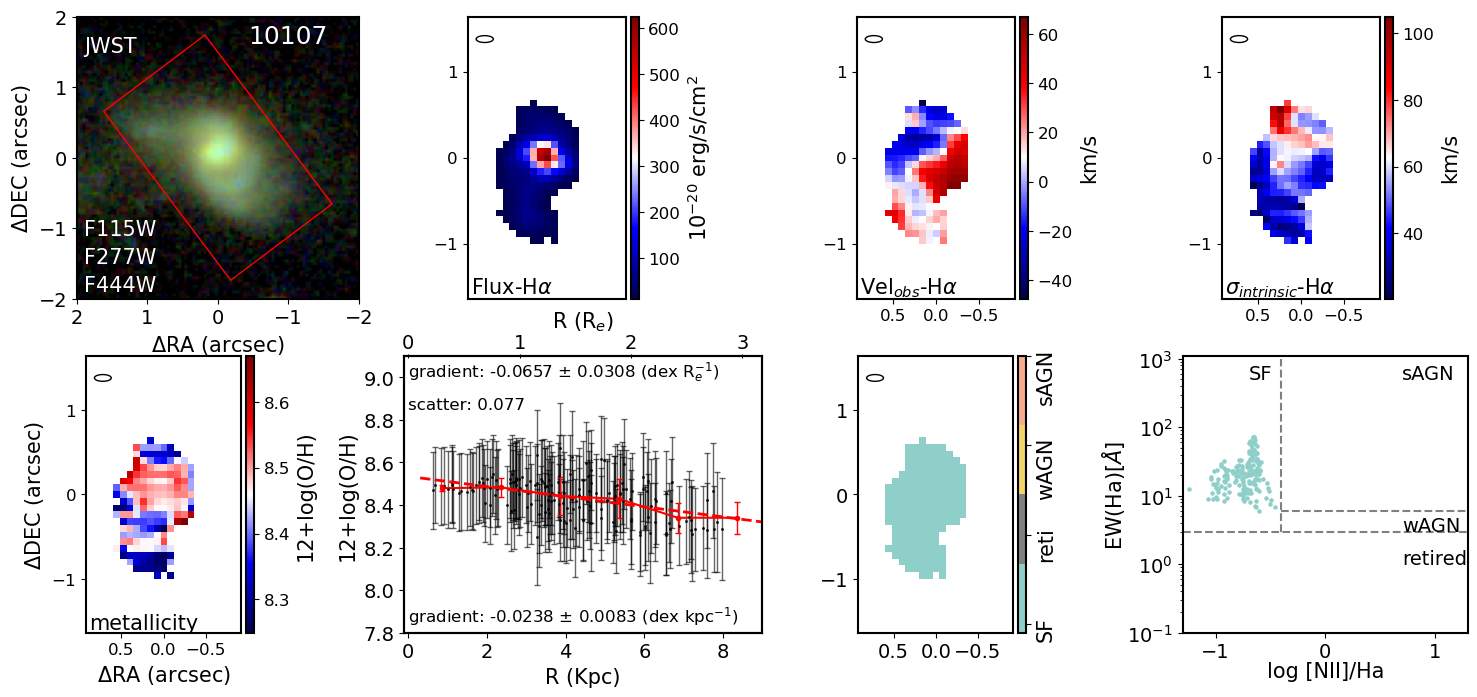}\\
    {\bf Figure A1.} continued\\
\end{figure*}
\begin{figure*}
\centering
    \ContinuedFloat
    \includegraphics[width=0.8\textwidth,clip,trim={0 0 0 0}]{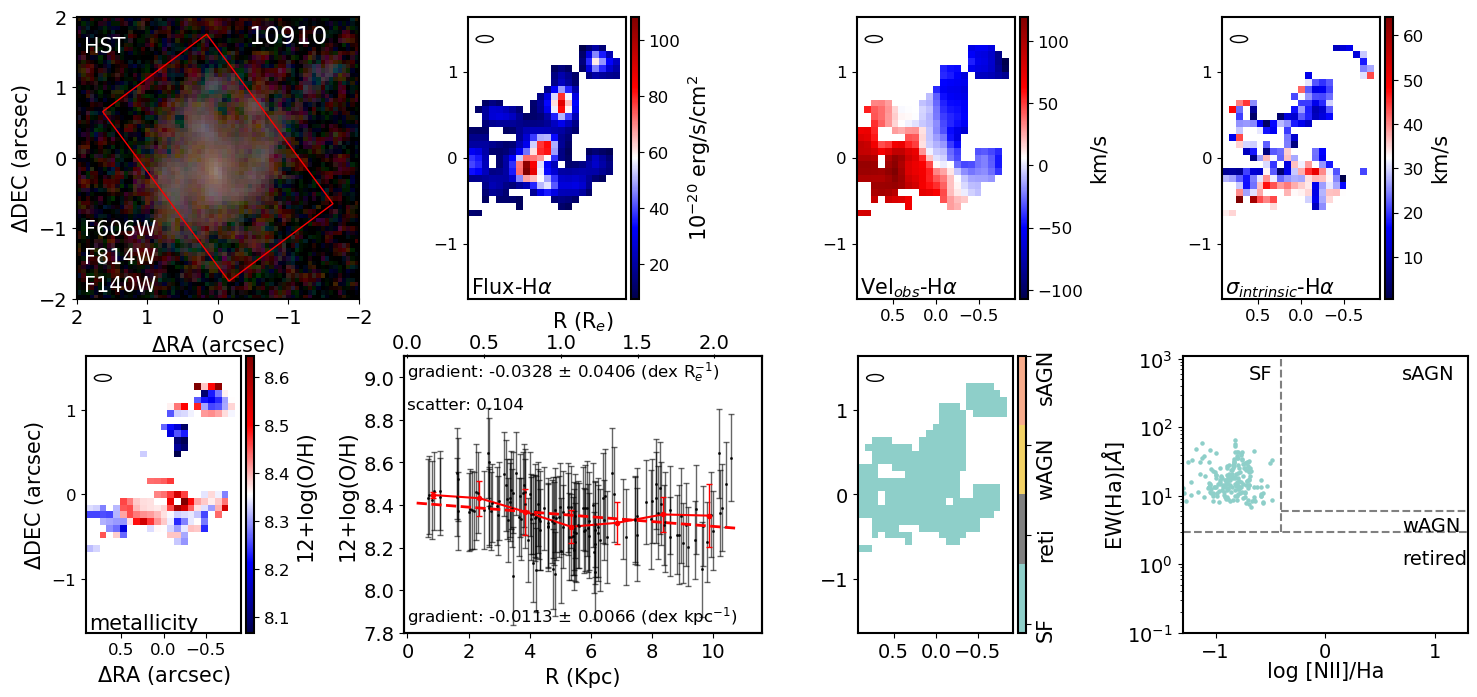}\\
    \includegraphics[width=0.8\textwidth,clip,trim={0 0 0 0}]{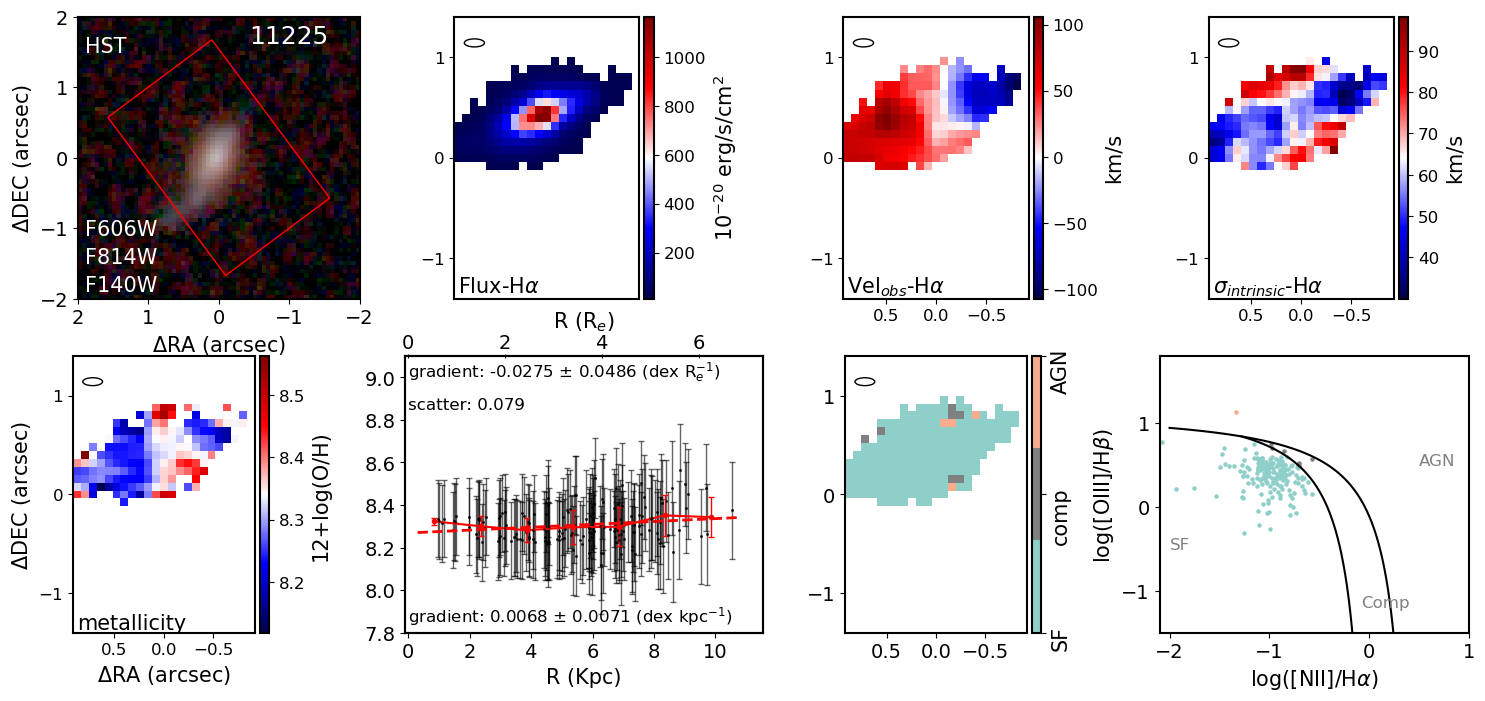}\\
    \includegraphics[width=0.8\textwidth,clip,trim={0 0 0 0}]{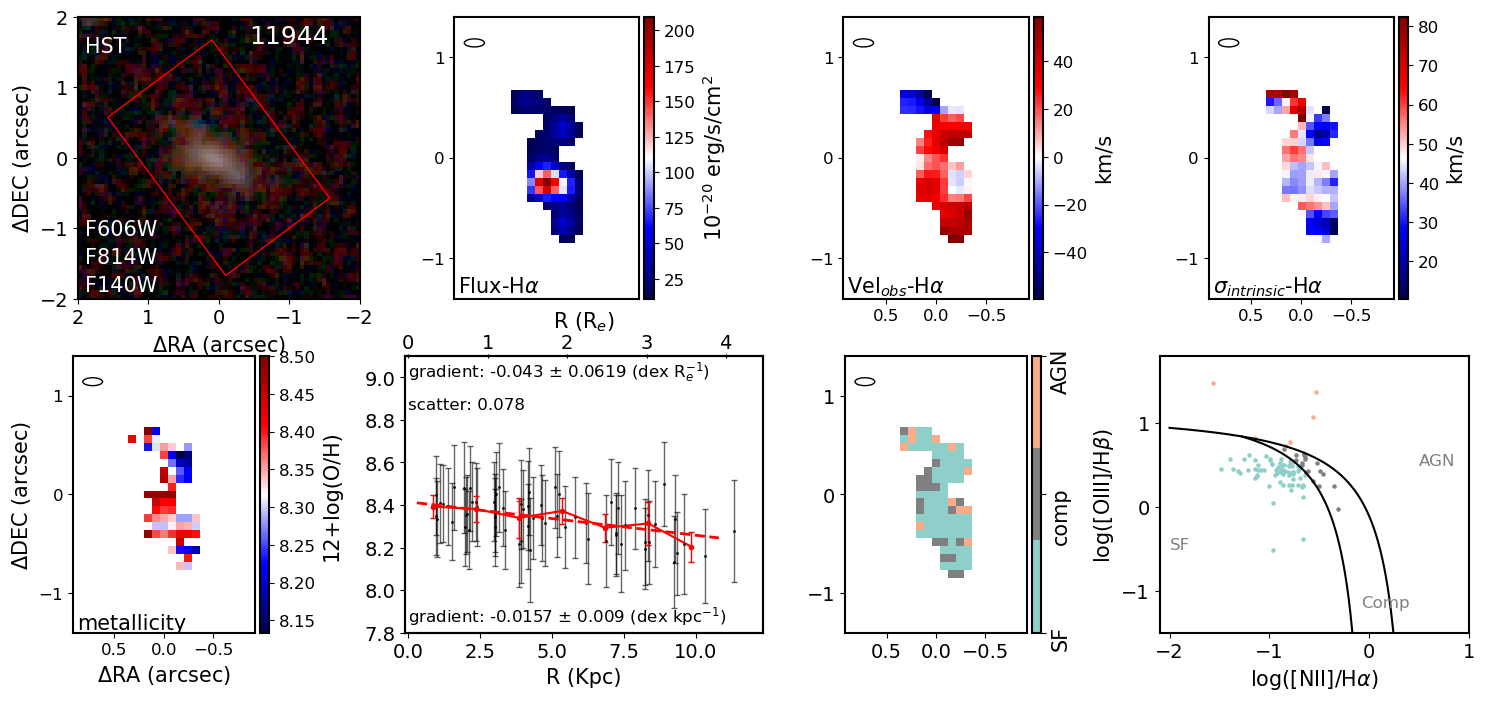}\\
    {\bf Figure A1.} continued\\
\end{figure*}
\begin{figure*}
\centering
    \ContinuedFloat
    \includegraphics[width=0.8\textwidth,clip,trim={0 0 0 0}]{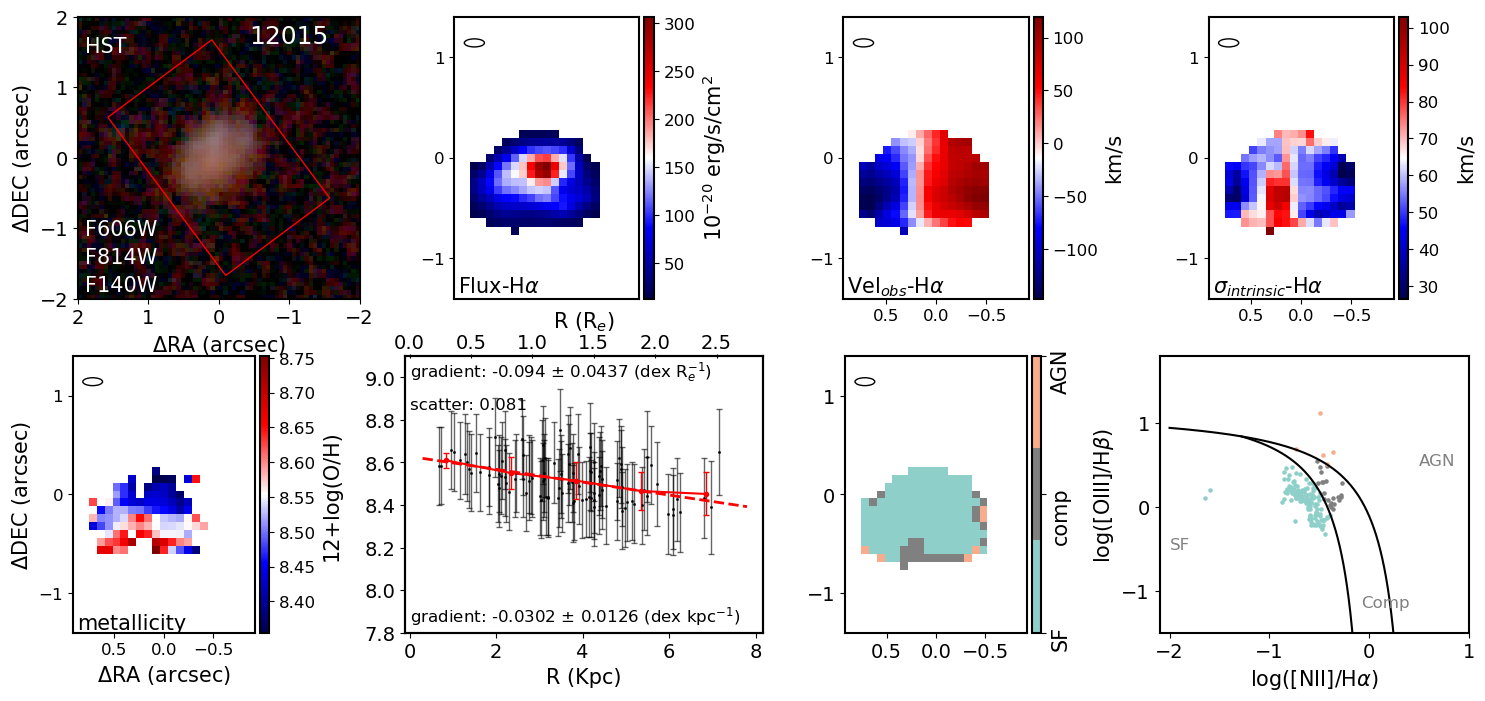}\\
    \includegraphics[width=0.8\textwidth,clip,trim={0 0 0 0}]{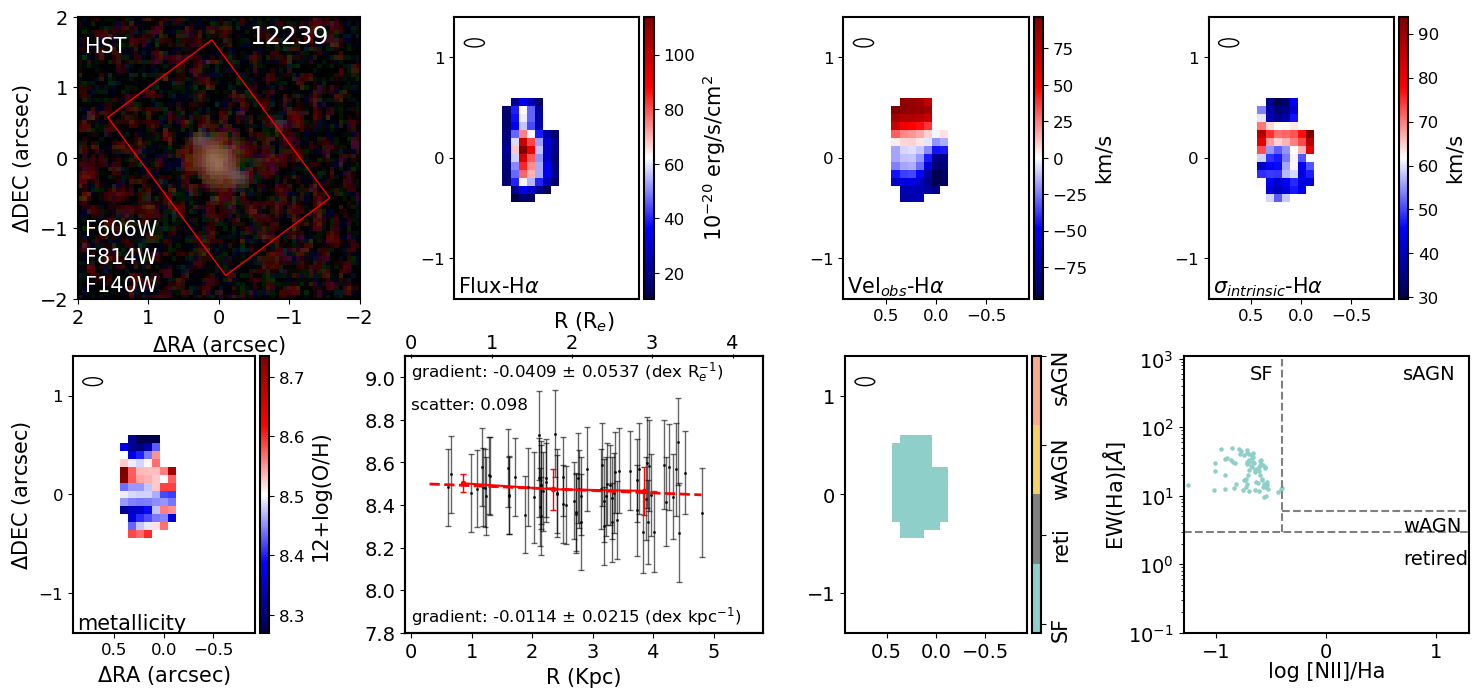}\\
    \includegraphics[width=0.8\textwidth,clip,trim={0 0 0 0}]{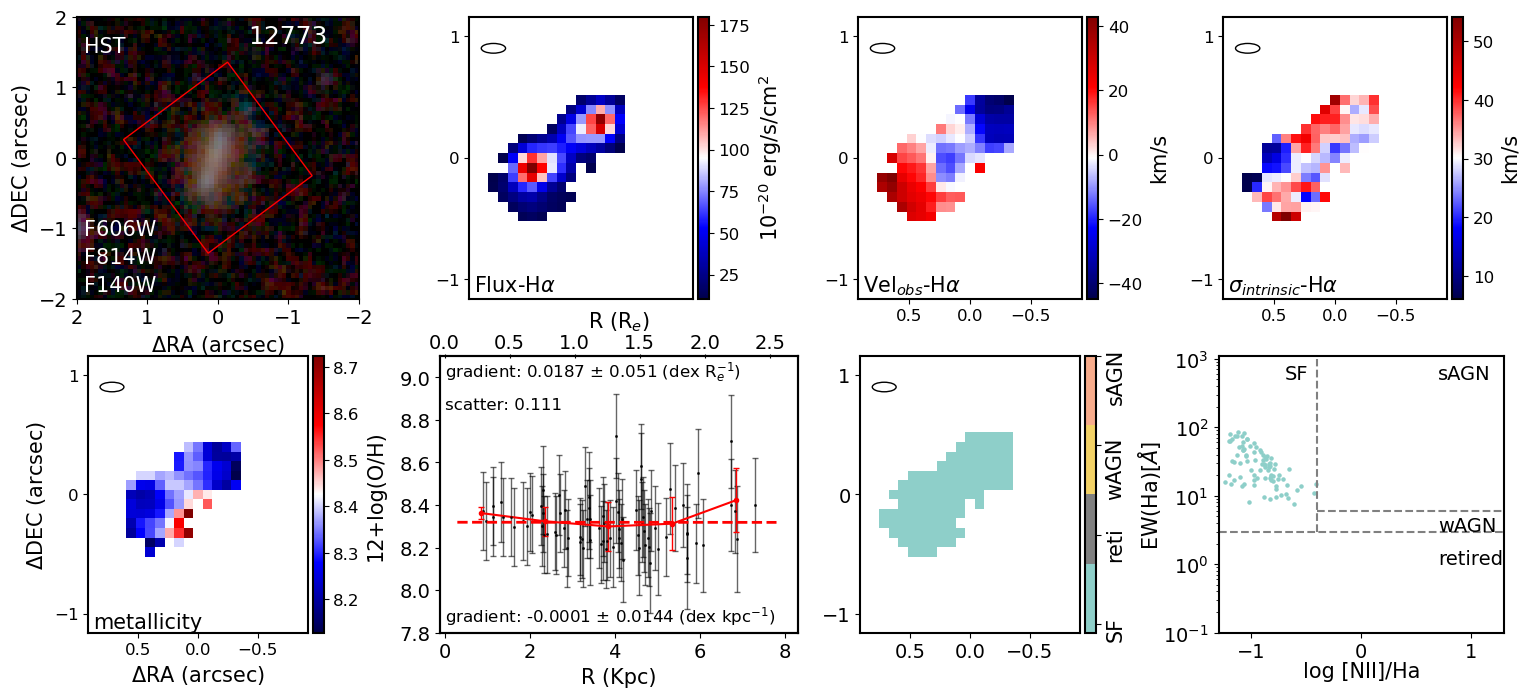}\\
    {\bf Figure A1.} continued\\
\end{figure*}
\begin{figure*}
\centering
    \ContinuedFloat
    \includegraphics[width=0.8\textwidth,clip,trim={0 0 0 0}]{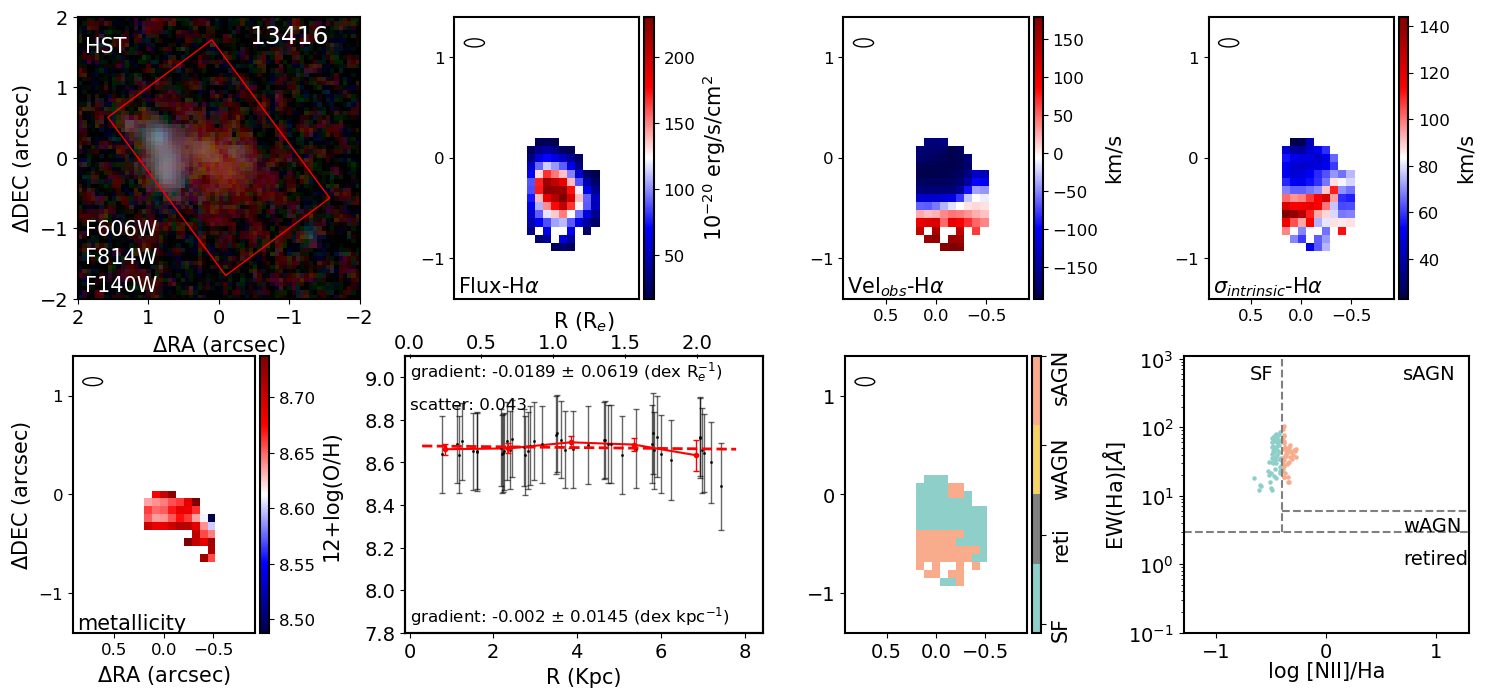}\\
    \includegraphics[width=0.8\textwidth,clip,trim={0 0 0 0}]{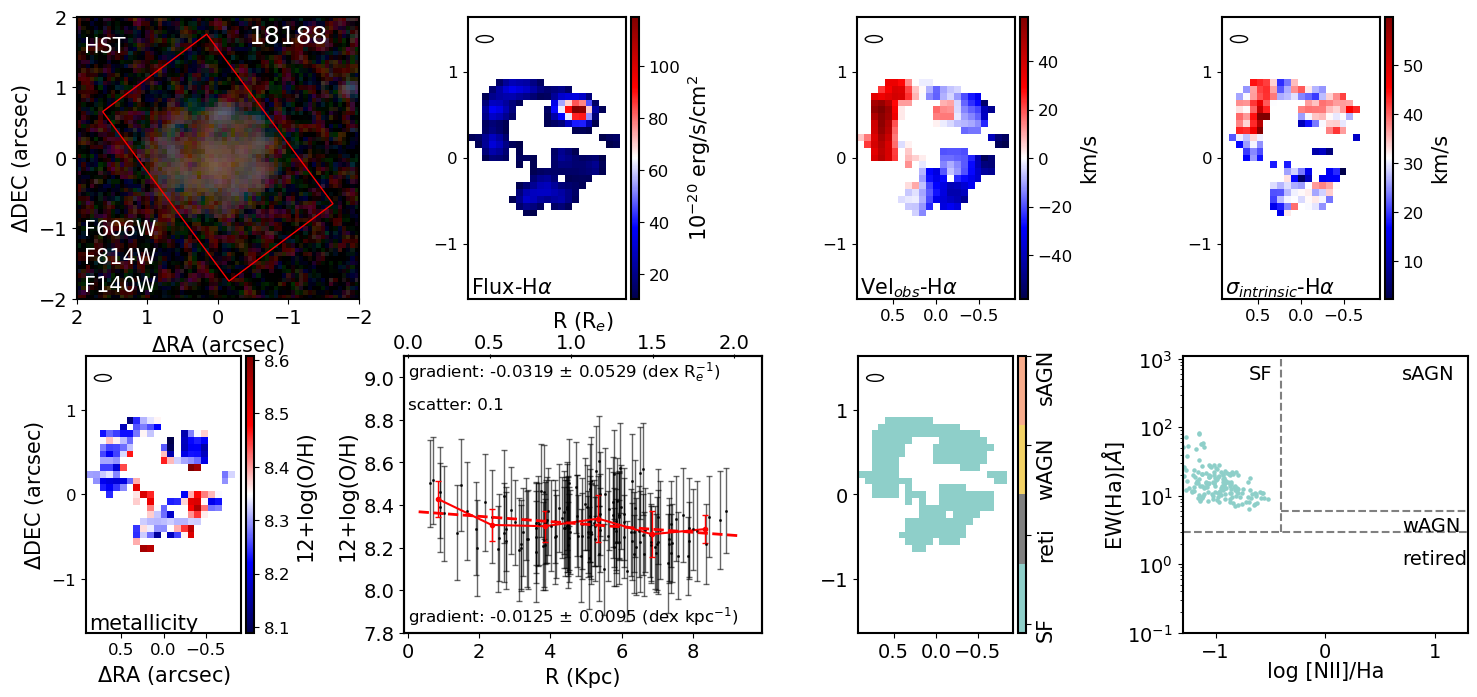}\\
    \includegraphics[width=0.8\textwidth,clip,trim={0 0 0 0}]{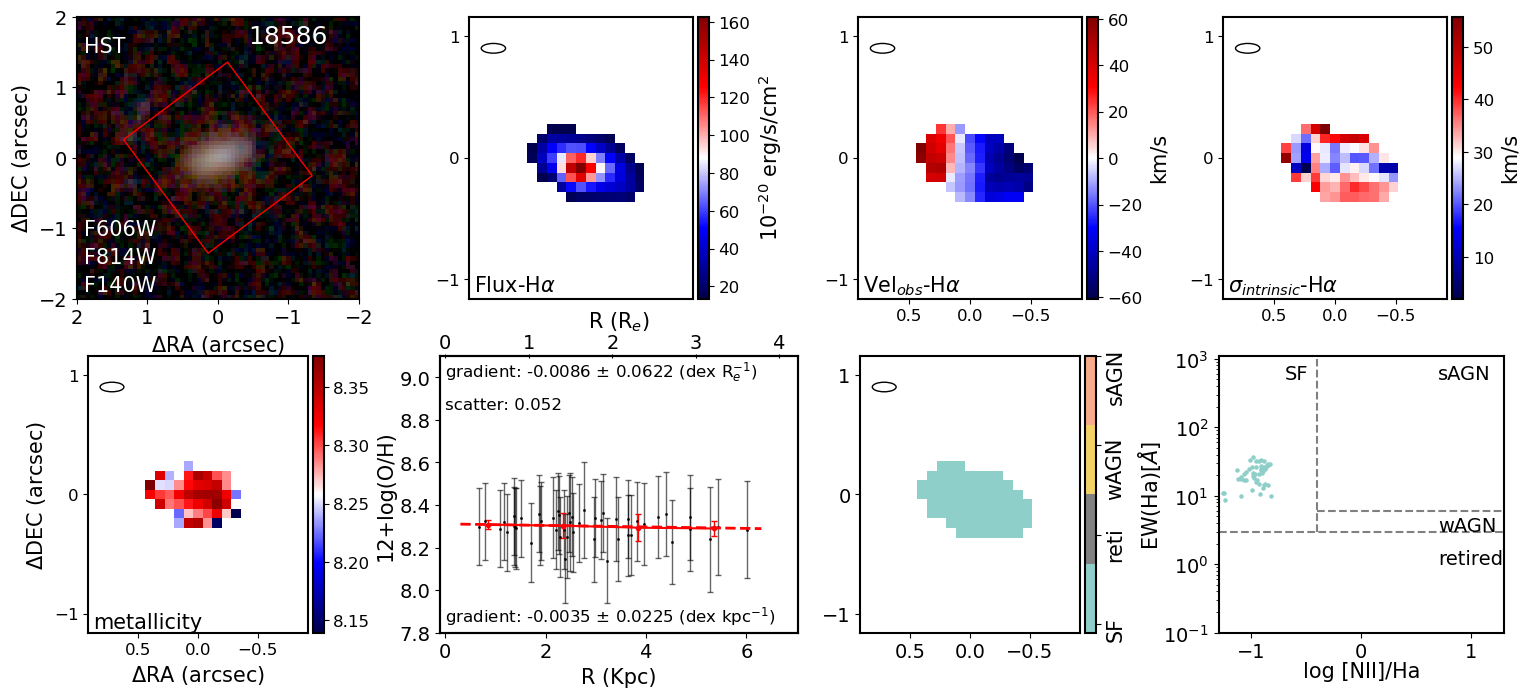}\\
    {\bf Figure A1.} continued\\
\end{figure*}
\begin{figure*}
\centering
    \ContinuedFloat

    \includegraphics[width=0.8\textwidth,clip,trim={0 0 0 0}]{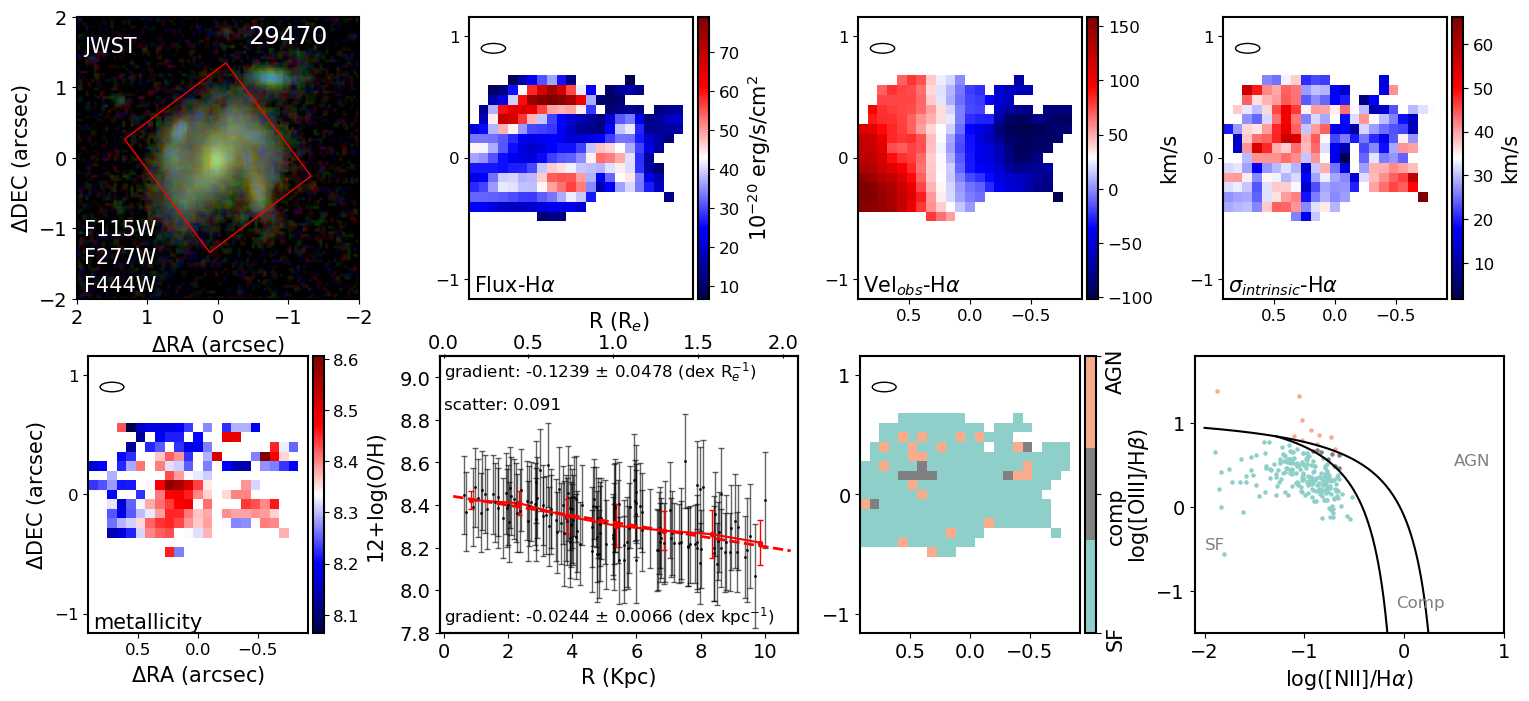}\\
    {\bf Figure A1.} continued\\
\end{figure*}

\bibliographystyle{aasjournal}
\bibliography{cita.bib}

\end{CJK*}
\end{document}